\documentclass[apj,twocolumn]{openjournal}
\usepackage{amsmath}
\usepackage{booktabs}
\usepackage{multirow}
\usepackage{color}
\usepackage{soul}
\usepackage[dvipsnames]{xcolor} 
\definecolor{ultra}{HTML}{0B6572}
\definecolor{udark}{HTML}{0A5681}
\usepackage[breaklinks,colorlinks,citecolor=ultra,urlcolor=udark,linkcolor=udark]{hyperref}
\renewcommand{\arraystretch}{1.2}  
\usepackage{listings}
\usepackage{orcidlink}
\allowdisplaybreaks

\begin{document}


\title{Instability and vertical eccentricity variation in global hydrodynamic disk simulations}

\author{\vspace{-40pt}Janosz W. Dewberry\orcidlink{0000-0001-9420-5194}$^{1}$}
\author{Henrik N. Latter$^{2}$}
\author{Gordon I. Ogilvie$^{2}$}
\author{Sebastien Fromang$^{3}$}
\affiliation{$^1$Department of Astronomy, University of Massachusetts Amherst, 710 N Pleasant St, Amherst, MA 01003, USA}
\affiliation{$^{2}$DAMTP, University of Cambridge, CMS, Wilberforce Road, Cambridge, CB3 0WA, UK}
\affiliation{$^{3}$Laboratoire des Sciences du Climat et de l'Environnement (LSCE-IPSL) CEA-CNRS-UVSQ-Université Paris Saclay 91190 Gif-sur-Yvette France}
\email{Corresponding author: jdewberry@umass.edu}

\begin{abstract}
Many dynamical interactions can induce eccentricities in astrophysical accretion disks. Disk eccentricities in turn seed a variety of instabilities, even in ideal hydrodynamics. We use 3D nonlinear simulations and $2+1$D linear calculations to characterize local and global instabilities in strongly distorted disks. On local scales, our simulations show the growth of parametrically excited inertial waves, which drive wave turbulence. The inertial waves' growth rates and localizations agree with the predictions of local theory. On global scales, we observe the growth of a separate family of low-frequency, vertically structured modes that compare favorably with eigenmodes computed from the linear theory of an eccentric background state. These low-frequency modes interact nonlinearly with the inertial wave turbulence driven by parametric instability, and they induce variation in eccentricity profiles that are initially uniform in the vertical direction. Extrapolating from our vertically local framework, we postulate that these secondary distortions may correspond to the corrugation of an initially planar eccentric disk. Our simulations demonstrate that strong disk eccentricities drive numerous dynamical phenomena even in a purely hydrodynamic, Newtonian framework.
\end{abstract}

\maketitle  

\section{Introduction}\label{sec:intro}
Many processes produce eccentric distortions in astrophysical accretion disks. Disk eccentricity driven by tidal interaction with a companion star provides a long accepted explanation for ``superhumps,'' luminosity variations with periods slightly longer than the orbital period of SU UMa dwarf novae \citep{Whitehurst1988,Lubow1991a,Lubow1991b,Ostriker1992,Hirose1993}. Tidal interactions should also produce eccentricities in the accretion flows of low-mass X-ray binary systems with extreme mass ratios \citep{Haswell2001,Patterson2005,Zurita2008,Ferreira2009}. The prograde precession of an eccentric circumstellar disk provides a probable explanation for the cyclical variation in intensity of Balmer lines seen in Be stars \citep{Okazaki1991,Papaloizou1992,Ogilvie2008}. Meanwhile, the tidal disruption of a star passing by a supermassive black hole can produce a highly eccentric accretion disk \citep{Gurzadian1979,Guillochon2014,Holoien2019}. Similarly, circumstellar disks can gain eccentricity during close encounters with another star \citep{Dai2015}. Protoplanetary disks should become eccentric due to tidal interactions with a binary companion \citep{Regaly2011}, or embedded gap-opening planets \citep{Goldreich1981,Teyssandier2016,Teyssandier2017}. The dynamic gravitational potential of a central binary also invariably produces eccentric circumbinary disks \citep[e.g.,][]{MacFadyen2008,Munoz2020,Munoz2023}.

The ubiquity of eccentric disks motivates careful consideration of their dynamics. Non-circular disks are intrinsically unstable \citep{Papaloizou2005a,Barker2014}, yet observations indicate that eccentricities persist. Is this persistence due solely to the presence of external driving, or can free eccentric distortions survive for long timescales? This paper looks beyond the properties of disk eccentricities themselves \citep[which have been extensively studied with both linear and nonlinear theory:][]{Tremaine2001,Statler2001,Ogilvie2001,Papaloizou2002,Goodchild2006,Ogilvie2008,Ogilvie2014,Ogilvie2019,Munoz2020}, to the secondary dynamical phenomena that they in turn induce. 

Distorted disks host a variety of such dynamical phenomena, providing one explanation for rapid variability in X-ray binary systems \citep{Kato2004,Ferreira2008,Dewberry2020a}, and more generically modifying the evolution of the magnetorotational instability \citep{Chan2018,Chan2022,Chan2024,Dewberry2020b}. Even in the absence of magnetic fields, purely hydrodynamic distorted disks are unstable to parametric instabilities. Parametric instability commonly manifests in asteroseismology as a resonant three-mode coupling, in which a pair of linearly damped ``children'' oscillation modes are destabilized by a linearly driven ``parent'' mode. In stars, the parent is usually an overstable acoustic (p-)mode, which transfers energy to internal gravity (g-)modes \citep[e.g.,][]{Dziembowski1985,Wu2002}. Distorted disks present a similar opportunity for parametric instability, but in this case the disk deformation is the parent mode, providing free energy to fuel the growth of inertial waves \citep{Goodman1993,Papaloizou2005a,Ogilvie2013,Barker2014}. The parametric instabilities of distorted disks are commonly studied from a local standpoint \citep{Ryu1994,Wienkers2018,Fairbairn2023,Held2024}, 
but they manifest in global disk simulations as well
\citep{Papaloizou2005b,Pierens2020,Pierens2021}. Characterizing the dynamics of eccentric disks in fully 3D, global models is a necessary step toward understanding their evolution in realistic astrophysical environments.

To further this characterization, we combine global unstratified hydrodynamic simulations of eccentric disks with a global linear theory for arbitrary non-axisymmetric background states. We use fully nonlinear eccentric disk solutions \citep{Barker2016,Ogilvie2019} as initial conditions in some simulations, and drive eccentricity from the outer boundary in others. Both approaches reduce shocks, such as those generated by the initialization of inexact linear eccentric modes in the simulations of \cite{Papaloizou2005b}. Our simulations therefore provide a clearer view of eccentricity's dynamical consequences. In addition to inertial wave turbulence and eccentricity decay, we observe the growth of coherent standing modes with very low frequencies. We identify these low-frequency modes as secondary, vertically structured distortions that modify the vertical profile of eccentricity.

Section \ref{sec:tbac} introduces and summarize the essential mechanism of the parametric instability, while in Section~\ref{sec:setup} we describe our numerical framework. We then present simulation results in Section~\ref{sec:results}, including a discussion in Section~\ref{sec:disk}, before concluding in Section~\ref{sec:conc}.

\section{Theoretical background}\label{sec:tbac}
At a basic level, parametric instability drives the growth of solutions $x=x(t)$ to the Hill equation written in the form \citep{Landau1969}
\begin{equation}\label{eq:LLpar}
    \dfrac{\text{d}^2 x}{\text{d} t^2}+\omega_0^2
    \left[1+p(t)\right]x=0.
\end{equation}
Here the ``pumping'' function $p(t)$ provides a small-amplitude, periodic deviation from the natural frequency $\omega_0$. This deviation imparts the strongest resonance when it has a frequency of nearly twice $\omega_0$, in which case growing solutions $x(t)$ can be found. 

Inertial waves in a circular disk are described locally by the unperturbed oscillator equation with $p=0$, and a natural frequency $\omega_0$ determined by the dispersion relation $\tilde{\omega}^2=k_z^2\kappa^2/k^2$. Here $k^2=k_r^2+k_z^2$, $k_r$ and $k_z$ are radial and vertical wavenumbers (referenced to cylindrical coordinates $r,\phi,z$), and $\tilde{\omega}=\omega-m\Omega$, where $\omega$ is the (inertial frame) wave frequency, $m$ is an azimuthal wavenumber, $\Omega$ is the fluid's rotation rate, and $\kappa$ is the horizontal epicyclic frequency. 

In the context of a deformed disk, $p(t)$ can be associated with a periodic perturbation of a given fluid element from its circular orbit by the warp or eccentricity. For low amplitude eccentricities (warps), this perturbation constitutes a horizontal (vertical) epicycle. Resonance then occurs when the natural frequency $\omega_0$ of the oscillations is given by half the local horizontal (vertical) epicyclic frequency, $\kappa/2$ $(\Omega_z/2)$. Since the spectrum of inertial waves is dense, vertical wavenumbers $k_r$ and $k_z$ can always be found that will lead to such resonance. 

From a Lagrangian viewpoint corotating with the fluid elements, the locally axisymmetric growing modes considered by \cite{Papaloizou2005a} and \cite{Barker2014} in nearly Keplerian disks then have frequencies $\sim\pm\Omega/2$, where $\Omega$ is understood to be the \emph{local} orbital frequency. In an Eulerian framework, the instability can be viewed as a three-mode coupling. In the linear regime, the ``parent'' eccentricity manifests as a mode with frequency $\omega_p\sim0$, azimuthal wavenumber $m_p\approx1$, and (in a cylindrical framework excluding vertical gravity) no vertical structure (i.e., $k_{z,p}=0$). The eccentric mode can then couple to secondary modes (denoted by subscripts $1$ and $2$) that satisfy 
$\omega_p\approx\omega_1\pm\omega_2,$ 
$m_p\approx m_1\pm m_2$, and 
$k_{z,p}=k_{z,1}\pm k_{z,2}.$ Choosing the negative sign, the parametric instability can be understood to excite pairs of inertial oscillations with the same vertical wavenumbers and frequencies ($+\Omega/2$ or $-\Omega/2$), and azimuthal wavenumbers differing by one \citep{Papaloizou2005a}.

\cite{Papaloizou2005a} derived a maximum growth rate of $3e\Omega/16$ for small-scale inertial waves localized to streamlines defined by Keplerian ellipses of constant eccentricity $e$, achieved when the waves have radial and vertical wavenumbers in a ratio of $k_r/k_z=\sqrt{3}.$  However, \cite{Papaloizou2005a} omitted vertical gravity and spatial gradients in eccentricity. In a fully 3D disk, the density scale height varies along eccentric streamlines, producing a laminar vertical oscillation \citep{Ogilvie2001}. Using a local model employing a modified shearing box, \cite{Barker2014} extended the stability analysis of \cite{Papaloizou2005a} from a uniformly eccentric disk without vertical structure to a disk model including both eccentricity gradients and vertical gravity. In so doing, the authors highlighted the relevance of these vertical oscillatory flows, which significantly enhance the growth rates of inertial waves excited by the parametric instability; with the inclusion of vertical gravity, \cite{Barker2014} found a maximum growth rate of $3e\Omega/4$ for a uniformly eccentric disk. 

Eccentricity gradients can further enhance the growth rate of the parametric instability. Following the local approach taken by \cite{Barker2014} but with vertical gravity and oscillatory flows suppressed, for small eccentricity we find a maximum growth rate 
\begin{equation}\label{eq:smax}
    s_{\rm max}=\frac{3}{16}|e + \lambda e'|\Omega
\end{equation}
in the limit $\tilde{k}_z\rightarrow\infty,$ and for $\tilde{k}_\xi^2\sim3(\tilde{k}_z^2-1/4)$. Here $e$ is the eccentricity, $\lambda=a(1-e^2)$ is the orbital semilatus rectum, $a$ is the orbital semimajor axis, and the prime denotes differentiation with respect to $\lambda$. Lastly $\tilde{k}_\xi$ and $\tilde{k}_z$ are dimensionless (quasi)radial and vertical wavenumbers given in units of the inverse of the local scale height.

The analyses of \cite{Papaloizou2005a} and \cite{Barker2014} examined the excitation of small-scale and incompressible inertial waves, and was thus scale-free. However, the instability may exhibit a global character, especially if the growing waves propagate significant radial distances before, or during, their saturation. For a start, growing disturbances may form as standing waves in radius (and thus as global eigenmodes), though they are still likely to localise around their resonant radius (they will disfavour non-resonant radii because of detuning). On the other hand, if there is a sufficient supply of noise, pairs of radially propagating waves might achieve large amplitudes via parametric instability as they pass through their resonant radii. They may either then break and saturate nearby, or continue travelling onwards to non-resonant radii, where they could slowly dissipate or interact nonlinearly with other (growing) waves \citep[cf.][]{Svanberg2022,Ogilvie2025}. If non-axisymmetric, the waves will be absorbed at their corotation resonance \citep{Li2003,Latter2009}, and this radius provides an upper bound on the radial domain the waves could traverse. 
\section{Numerical methods}\label{sec:setup}
We run nonlinear simulations with a uniform-mesh version of the code RAMSES \citep{Teyssier2002,Fromang2006,Faure2014}, which uses a high-order, conservative Godunov scheme to solve the magnetohydrodynamic equations.\footnote{The version of the code used here is freely available at \hyperlink{https://sourcesup.renater.fr/projects/dumses/}{https://sourcesup.renater.fr/projects/dumses/}
} In the absence of magnetic fields, these reduce to the hydrodynamic relations 
\begin{align}
    \dfrac{\partial \rho }{\partial t}
    +\nabla\cdot(\rho{\bf u})&=0,
\\
    \dfrac{\partial (\rho{\bf u} )}{\partial t}
    +\nabla\cdot(\rho {\bf u}{\bf u} + P{\bf I})
    &=-\rho\nabla\Phi,
\end{align}
where $\rho$ is the density, $\bf u$ is the fluid velocity, $P$ the gas pressure, and $\Phi$ is the gravitational potential of a central mass. We take the equation of state to be globally isothermal, with $P=c_s^2\rho$ and $c_s$ purely constant. Our simulations use HLLC Riemann solvers, second order reconstruction with MinMod slope limiters, and a Courant factor of 0.8.

We focus on behavior at the mid-plane with a cylindrical approximation, eschewing vertical gravity and density stratification to solve the equations on a fixed grid that is uniform in the coordinates $(r,\phi,z)$. Under this approximation, the gravitational potential is simply $\Phi=-GM/r,$ where $M$ is the mass of the central body, and $G$ is the gravitational constant. In addition to significantly reducing numerical expense, this simplified, radially global but vertically local framework provides a context for a more controlled experiment (see \autoref{sec:diag}). 

We take $r_0$ and $\Omega_0^{-1}=[r_0^3/(GM)]^{1/2}$ as our units in space and time, respectively, and quote velocities in units of $r_0\Omega_0$ throughout. The isothermal sound speed $c_s=0.05r_0\Omega_0$ in all cases. With a constant sound speed the scale height varies as $H/r=0.05\sqrt{r/r_0}$, meaning that the effective vertical resolution increases with radius. Owing to the relatively small ratios of $r_1/r_0$ considered, however, this change in vertical resolution is minimal. Tables \ref{tab:2Dsim} and \ref{tab:3Dsim} summarize salient details for our 2D and 3D simulations (respectively).

\subsection{Initial conditions}\label{sec:ic}
\begin{figure}
  \includegraphics[width=\columnwidth]{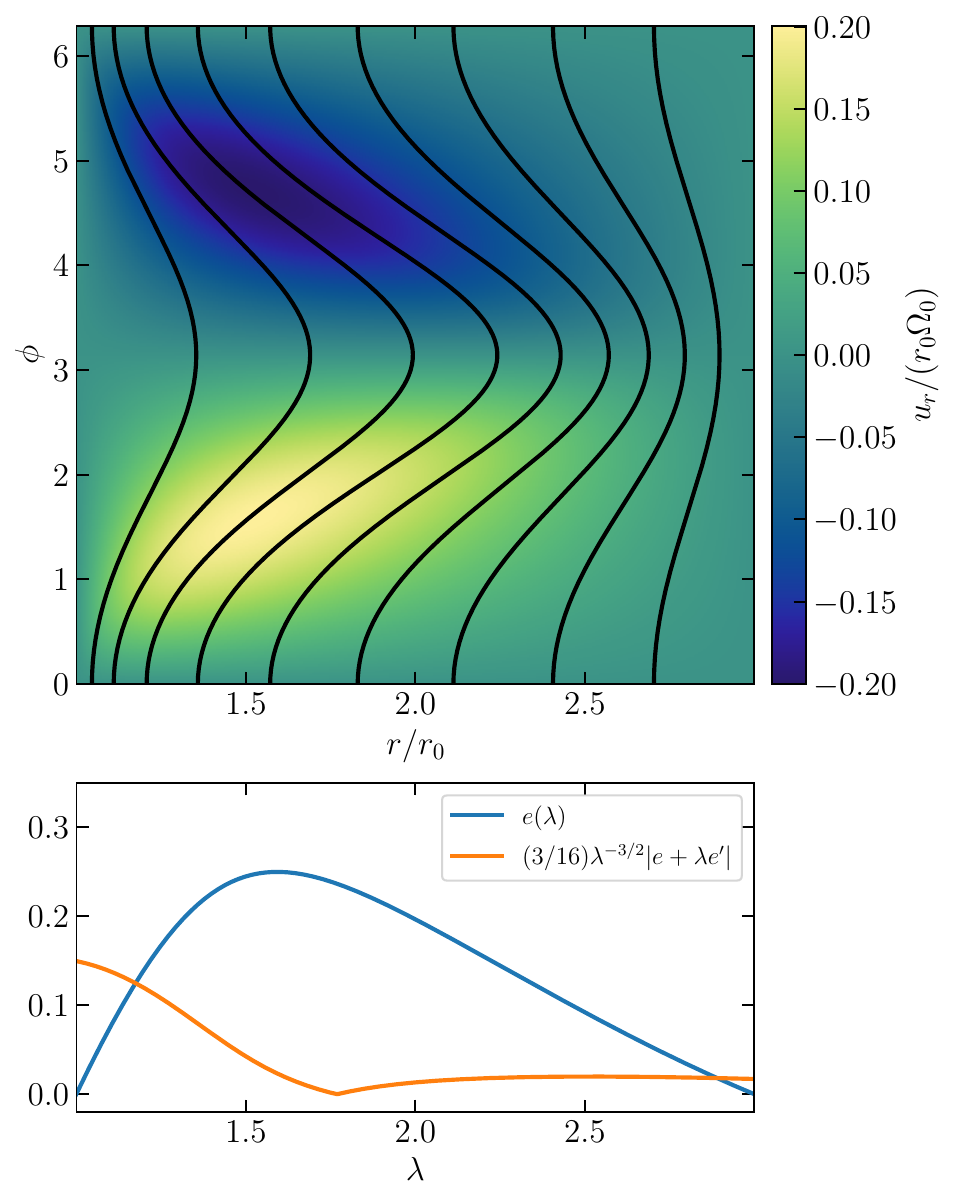}
    \caption{Top: unfolded plot showing the radial velocity field associated with an eccentric disk solution to a non-linear secular theory, calculated to achieve a maximum eccentricity of $0.25$ in the middle of the radial domain ($r\in[r_0,3r_0],$ $c_s=0.05r_0\Omega_0$). The streamlines superimposed as black lines illustrate the disk's eccentricity (circular streamlines would be vertical on the unfolded polar plot). Bottom: corresponding eccentricity profile (blue) and maximum growth rate (in units of $\Omega_0$) for the local parametric instability (orange), both plotted as a function of semilatus rectum $\lambda$.}
\label{fig:A25vr}
\end{figure}

We introduce eccentricity into our disk simulations using two methods. First, we build upon the approach of \cite{Papaloizou2005b} by initializing both 2D and 3D simulations with eigenmode solutions of the nonlinear secular theory, as computed in \cite{Barker2016} and \cite{Ogilvie2019}. These solutions have radially varying profiles in eccentricity that vanish at the circular boundaries, and achieve a maximum eccentricity in the middle of the domain. \autoref{fig:A25vr} shows an example of such a disk solution, computed with a maximum eccentricity of $0.25$. Large (supersonic) radial velocities result from the relatively large eccentricity. 

In some simulations we also drive eccentricity from the outer radial boundary $r_1$, using a boundary condition employed by \cite{Dewberry2020a,Dewberry2020b} in simulations of relativistic disks (see \autoref{sec:bc}). Since the disk simulations described in this work do not possess an innermost stable circular orbit or plunging region, the eccentric wave generated at $r_1$ reflects off the inner boundary $r_0$ and forms a standing mode very similar to the one pictured in \autoref{fig:A25vr}. We construct these disks with forced eccentricity first in 2D simulations. Once these have been evolved to a quasi-steady state, we use the end-state eccentric disks as initial conditions for 3D simulations. 

In simulations with both free and forced eccentricity, we seed instability by imposing white noise, subsonic velocity perturbations.

\begin{table*}
\caption{\normalfont
Table of 2D simulations, listing domain size, resolution, initial condition type, initial maximum value of eccentricity, ending maximum value of eccentricity, runtime, predicted precession period, and measured precession period. For simulations with eccentricity driven from the outer boundary, $\max[e]_i$ gives the maximum eccentricity of the free mode used to generate an outer boundary condition (which is generally larger than the resulting maximum eccentricity in the domain). 
}\label{tab:2Dsim}
\centering
\begin{tabular}{lcccccccr} 
    \hline
    Label &
    $r_1/r_0$ &
    $N_r\times N_{\phi}$ &
    IC & 
    $\max[e]_i$ & 
    $\max[e]_f$ & 
    $T_{max}\Omega_0$ &
    $P_t\Omega_0$ & 
    $P_0\Omega_0$ \\
    \hline
    A05r32D & $3$ & $512\times 512$ & Free mode & 0.05 & 0.05 & 2500 & 720 & 735 \\
    A10r32D & $3$ & $512\times 512$ & Free mode & 0.10 & 0.09 & 2500 & 700 & 701 \\
    A15r32D & $3$ & $512\times 512$ & Free mode & 0.15 & 0.12 & 2500 & 667 & 652 \\
    A20r32D & $3$ & $512\times 512$ & Free mode & 0.20 & 0.14 & 2500 & 622 & 608 \\
    {A25r32D}  & $3$ & $512\times 512$ & {Free mode} & $0.25$ & $0.16$ & $2500$ & $563$ & $537$ \\
    A25r32Dhr & $3$ & $1024\times 512$ & Free mode & 0.25 & 0.21 & 1250 & 563 & 554 \\
    A25r32DHR & $3$ & $1024\times 1024$ & Free mode & 0.25 & 0.19 & 2500 & 563 & 553 \\
    A30r32D & $3$ & $512\times 512$ & Free mode & 0.30 & 0.17 & 2500 & 489 & 490 \\
    A25r52D & $5$ & $1024\times 512$ & Free mode & 0.25 & 0.21 & $2500$ & $1461$ & $1419$ \\
    A35r52D & $5$ & $1024\times 512$ & Free mode & 0.35 & 0.27 & $2500$ & $1325$ & $1268$ \\    
    Af15r32D & $3$ & $512\times 512$ & Driven ecc & $0.15$ & 0.10 & $12500$ & -- & -- \\
    Af20r32D & $3$ & $512\times 512$ & Driven ecc & $0.20$ & 0.12 & $12500$ & -- & -- \\
    Af25r32D & $3$ & $512\times 512$ & Driven ecc & $0.25$ & 0.13 & $12500$ & -- & -- \\
    Af30r32D & $3$ & $512\times 512$ & Driven ecc & $0.30$ & 0.15 & $12500$ & -- & -- \\
    Af35r32D & $3$ & $512\times 512$ & Driven ecc & $0.35$ & 0.18 & $12500$ & -- & -- \\
 \end{tabular}
\end{table*}

\begin{table*}
\centering
\caption{\normalfont
Table of 3D simulations, listing domain size, resolution, initial condition type, vertical boundary condition, initial maximum value of eccentricity, ending maximum value of eccentricity, runtime, predicted precession period, measured precession period, and growth rate. Runs A25r3D2H and A25r3Dp2H have a vertical extent $z\in[-2H_0,2H_0],$ and all the others have $z\in[-H_0,H_0],$ where $H_0=c_s/\Omega_0$.
}\label{tab:3Dsim}
\begin{tabular}{lcccccccccr} 
    \hline
    Label &
    $r_1/r_0$ &
    $N_r\times N_{\phi}\times N_z$ &
    IC &
    Upper BC &
    $\max[e]_i$ & 
    $\max[e]_f$ & 
    $T_{max}\Omega_0$ &
    $P_t\Omega_0$ & 
    $P_0\Omega_0$ & 
    $s/\Omega_0$ \\
    \hline
        A05r3D  & $3$ & $512\times 512\times 32$ & Free & Rigid & 0.05 & 0.05 & $2500$ & $720$ & $730$ & $--$ \\
        A15r3D  & $3$ & $512\times 512\times 32$ & Free & Rigid & 0.15 & 0.11 & $2500$ & $667$ & $658$ & $3\times 10^{-2}$ \\
        A25r3D  & $3$ & $512\times 512\times 32$ & Free & Rigid & $0.25$ & $0.15$ & $2500$ & $563$ & $539$ & $ 7\times 10^{-2}$ \\
        A25r3D2H & $ 3$ & $512\times 512\times 64$ & Free & Rigid & $0.25$ & 0.12 & $2500$ & 563 & 549 & $6\times10^{-2}$ \\
        A25r3Dhr  & $3$ & $1024\times 512\times 32$ & Free & Rigid & $0.25$ & $0.18$ & $1250$ & 563 & 573 & $7\times 10^{-2}$ \\
        A25r3DHR & $ 3$ & $1024\times 1024\times 64$ & Free & Rigid & $0.25$ & $0.16$ & $2500$ & 563 & 560 & $7\times10^{-2}$ \\
        A25r53D & $5$ & $1024\times 512\times 32$ & Free & Rigid & 0.25 & 0.21 & 2500 & 1461 & $141$ & $3\times 10^{-2}$ \\
        A35r53D & $5$ & $1024\times 512\times 32$ & Free & Rigid & 0.35 & 0.27 & 2500 & 1325 & $1230$ & $4\times 10^{-2}$ \\
        Af25r3D & $3$ & $512\times 512\times 32$ & Driven & Rigid & $0.13$ & $0.13$ & $2500$ & -- & $564$  & $3\times 10^{-3}$ \\
        Af35r3D & $3$ & $512\times 512\times 32$ & Driven & Rigid & 0.18 & 0.17 & $2500$ & -- & $402$  & $9\times 10^{-3}$ \\
        A10r3Dp & $ 3$ & $512\times 512\times 32$ & Free & Periodic & $0.10$ & 0.09 & $1250$ & 700 & 701 & -- \\
        A15r3Dp & $ 3$ & $512\times 512\times 32$ & Free & Periodic & $0.15$ & 0.13 & $1250$ & 667 & 654 & $1\times10^{-2}$\\
        A20r3Dp & $ 3$ & $512\times 512\times 32$ & Free & Periodic & $0.20$ & 0.16 & $1250$ & 622 & 605 & $2\times10^{-2}$\\
        A25r3Dp & $ 3$ & $512\times 512\times 32$ & Free & Periodic & $0.25$ & 0.16 & $2500$ & 563 & 538 & $3\times10^{-2}$\\
        A25r3DpHR & $ 3$ & $1024\times 1024\times 64$ & Free & Periodic & $0.25$ & 0.21 & $1250$ & 563 & 555 & $7\times10^{-2}$ \\
        A25r3Dp2H & $ 3$ & $512\times 512\times 64$ & Free & Periodic & $0.25$ & 0.16 & $2500$ & 563 & 535 & $6\times10^{-2}$ \\
        A30r3Dp & $ 3$ & $512\times 512\times 32$ & Free & Periodic & $0.30$ & 0.21 & $1250$ & 489 & 475 & $5\times10^{-2}$\\
        Af20r3Dp & $ 3$ & $512\times 512\times 32$ & Driven & Periodic & 0.12 & 0.11 & $1250$ & -- & 621 & -- \\
        Af25r3Dp & $ 3$ & $512\times 512\times 32$ & Driven & Periodic & 0.13 & 0.13 & $1250$ & -- & 562 & $3\times10^{-3}$\\
        Af30r3Dp & $ 3$ & $512\times 512\times 32$ & Driven & Periodic & 0.15 & 0.15 & $1250$ & -- & 490 & $1\times10^{-2}$\\
        Af35r3Dp & $ 3$ & $512\times 512\times 32$ & Driven & Periodic & 0.18 & 0.18 & $1250$ & -- & 399 & $3\times10^{-2}$\\
    \hline
    \\
 \end{tabular}
\end{table*}

\subsection{Boundary conditions}\label{sec:bc}
We implement periodic boundary conditions in $\phi$, which spans the full azimuthal range in all our runs. In simulations initialized with non-linear eccentric eigenmodes, we employ quasi-rigid walls at both the inner and outer radial boundaries ($r_0$ and $r_1$). We match the density in the ghost cells to the nearest cell in the active domain, set the radial and vertical velocities to zero, and extrapolate the azimuthal velocity to the Keplerian rotation velocity, $r\Omega_K(r)\propto r^{-1/2}.$ The reflection provided by these radial boundary conditions is not realistic, and we note that it likely has at least some effect on the dynamics considered here (particularly at the inner boundary). 

To drive eccentricity from the outer boundary, we replace the zero-gradient condition imposed on density at $r_1$ with a non-axisymmetric density profile taken from a secular eccentric disk solution. At each time-step we set the density in the ghost cells to $\rho(r_1,\phi,z,t)=\rho_0(r_1,\phi+\omega_Pt),$ where $\rho_0$ and $\omega_P<0$ are the density and the precession frequency calculated from the secular theory. This produces a non-axisymmetric pressure gradient that forces the inward propagation of eccentricity. Note that the maximum eccentricity values of the solutions used to generate outer boundary conditions (listed in \autoref{tab:3Dsim}) do not exactly equal the maximum eccentricities that result in the simulation domain. 

\begin{figure*}
    \centering
    \includegraphics[width=\textwidth]{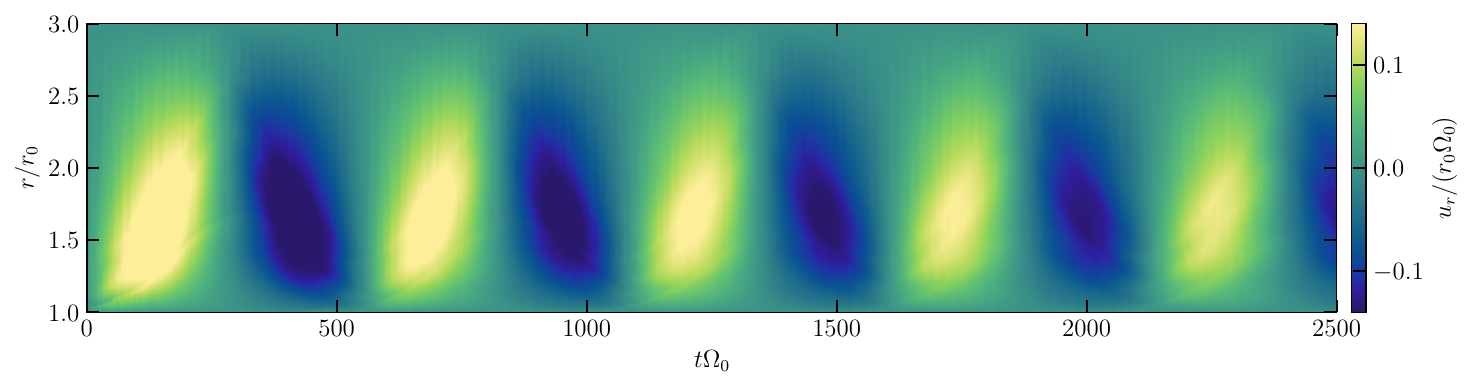}
    \caption{Spacetime diagram showing radial velocity (evaluated at a fixed $\phi=0$) as a function of radius (\textit{y}-axis) and time (\textit{x}-axis) for a 2D simulation (A25r32D) initialized with a free eccentric mode with a maximum eccentricity of $0.25$ (and a radial domain with $r_1/r_0=3$).}\label{fig:A25r32D_spct}
\end{figure*}

With a cylindrical approximation, it is most natural to impose periodic boundary conditions at the vertical boundaries. However, combined with this choice, at least one of the instabilities in our simulations generates mean vertical flows. These ``elevator'' flows \citep{Calzavarini06, Lohse24} likely do not represent a realistic outcome for disks with vertical gravity, and in our simulations they saturate the parametric instability prematurely by throwing the small-scale inertial oscillations out of resonance. For this reason, we also run simulations in which we set the vertical mass flux to zero in the vertical boundaries, and use zero-gradient boundary conditions for the rest of the hydrodynamic variables. Aside from halting the formation of the elevator flows, this new vertical boundary condition forces the excited inertial waves and global oscillations to become standing (rather than vertically traveling) waves.

\subsection{Diagnostics}\label{sec:diag}
We define the volume average of a given quantity $X$ as 
$\langle X\rangle_V = \int_V X dV/\int_VdV$, and refer to analogously defined azimuthal and vertical averages $\langle X\rangle_{\phi}$ and $\langle X\rangle_z$. The volume-integrated kinetic energy in radial motions is dominated by the eccentric modes in our simulations, and therefore tracks both the growth and the decay of the disk deformations. Values of eccentricity itself provide an additional quantification of distortion. We compute orbital parameters for each grid cell as described in \citet{Lynch2023}.

Vertical kinetic energy provides a useful tracer for eccentricity-driven instability, particularly in cylindrical simulations. Vertically local frameworks do not capture upward and downward motions caused by a variation in vertical gravity around elliptical streamlines in truly 3D eccentric disks. Vertical kinetic energy can therefore be dissociated from the essentially 2D eccentric modes considered in our simulations, and identified instead with motions excited by the eccentricity. We isolate these motions further by considering fluctuations of the form $X-\langle X\rangle_z$ for a given quantity $X$. 

Finally, quantitative analysis of the frequencies of the oscillations excited in our simulations requires calculation of the power spectral density (PSD) given by $P(\omega)=|\mathcal{F}(h)|^2$ where $\mathcal{F}(h)$ is the temporal Fourier transform of the signal $h(t)$. We primarily consider radially dependent PSDs $P(r,\omega)$ computed from radius and time dependent spacetime data.

\subsection{Linear calculations}
To better understand the dynamics that manifest in our simulations, we also compute the linear modes associated with our strongly eccentric disks. In a frame rotating with angular velocity $\boldsymbol{\Omega}_p=\omega_p\hat{\bf e}_z$ (chosen to follow the eccentric disk's precession), the linearized equations governing isothermal velocity and (normalized) density perturbations $\bf v$ and $h=\delta\rho/\rho_0$ to a given background state ${\bf u}_0,\rho_0$ are
\begin{align}\label{eq:vecl1}
    \frac{\partial {\bf v}}{\partial t}
    +{\bf u}_0\cdot\nabla {\bf v}
    +{\bf v}\cdot\nabla{\bf  u}_0
    +2\boldsymbol{\Omega}_p\times {\bf v}
    &=
    -c_s^2\nabla h,
\\\label{eq:vecl2}
    \frac{\partial }{\partial t}(\rho_0h)
    +\nabla\cdot[\rho_0({\bf v}+{\bf u}_0h)]
    &=0.
\end{align}
Appendix \ref{app:lin} describes our numerical method for computing eigenmode solutions ${\bf v},h\propto\exp[\text{i}(k_zz-\omega t)]$ to Equations \eqref{eq:vecl1} and \eqref{eq:vecl2}, which are non-separable in $r$ and $\phi$ when the background disk is eccentric.

\begin{figure*}
    \centering
    \includegraphics[width=0.95\textwidth]{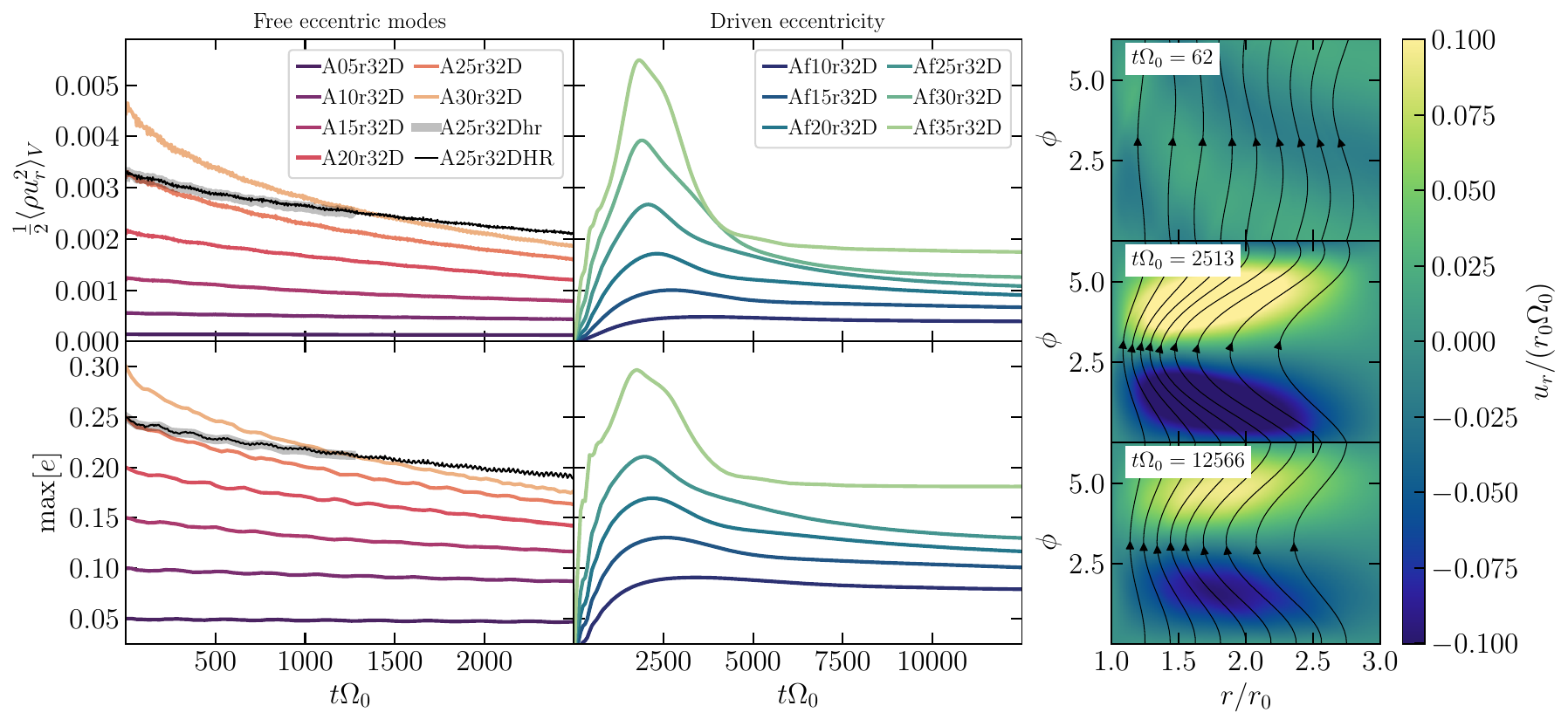}
    \caption{Left panels: radial kinetic energy (top) and maximum eccentricity values (bottom) for 2D simulations initialized with free eccentric modes (left) and driven eccentric distortions (right). Right colormaps: snapshots showing radial velocity for a simulation with driven eccentricity. The $\phi$-profile of density enforced in the ghost cells at the outer boundary is taken from a non-linear eccentric eigenmode with maximum eccentricity $\max[e]=0.25$.}\label{fig:2DrKE}
\end{figure*}

\section{Simulation results}\label{sec:results}

\subsection{2D simulations }\label{sec:nPI2D}
Our purely 2D simulations of eccentric disks serve to establish a point of comparison, and to produce initial conditions for our 3D simulations with forced eccentricity. We focus on simulation runtimes comparable to our 3D runs, since the long-term evolution of eccentric modes in 2D simulations has already been quantified by \cite{Barker2016}. In Keplerian thin disks the parametric instability affects the eccentricity on a faster timescale than, for example, the thin-disk manifestation of the Papaloizou--Pringle instability \citep{Papaloizou1984,Goldreich1986,Narayan1987,Godon1998}. Simulations run over only a few hundred inner orbital periods $T_\text{orb}=2\pi\Omega_0^{-1}$ therefore provide a sufficient comparison for 3D runs.

Our 2D simulations initialized with non-linear solutions to the secular theory largely follow the evolution observed by \cite{Barker2016}. The spacetime diagram in \autoref{fig:A25r32D_spct} demonstrates typical evolution. The colormap indicates radial velocity as a function of radius (\textit{y}-axis) and time (\textit{x}-axis), as measured at a fixed $\phi$ in the inertial frame of the simulation. The periodic variation in $u_r$ tracks the retrograde precession of the eccentric mode. In all of our 2D simulations, we measure an initial precession period (of $\sim 555\Omega_0^{-1}$ for the case shown in \autoref{fig:A25r32D_spct}) that agrees with the predictions of the secular theory to within $\sim 1-2\%$ (see Table \ref{tab:2Dsim}). 

As demonstrated by a gradual decrease in color saturation in \autoref{fig:A25r32D_spct}, the velocity variations associated with the eccentric mode decay in amplitude over time. This corresponds to the decay in radial kinetic energy and maximum eccentricity shown in \autoref{fig:2DrKE} (left), and is accompanied by a lengthening precession period. For low values of maximum eccentricity ($\max[e]\lesssim0.2$ for $r_1/r_0=2,$ $\max[e]\lesssim0.3$ for $r_1/r_0=3$) this decay is due largely to damping by numerical diffusion on the grid scale, which is exacerbated by the more densely packed streamlines. We find no clear evidence of Papaloizou-Pringle instability over the relatively short simulation run-times considered here. However, for sufficiently large eccentricity or narrow radial extents, we find \citep[like ][]{Barker2016} that shocks due to $\mathcal{O}(c_s^2)$ errors in the secular theory and interaction with the boundaries can strongly modify the disk eccentricity profile within a single precession period. Such shocks obscure the instabilities under consideration, so we focus on simulations with $r_1/r_0\geq3-5$, and maximum eccentricities of $0.05-0.35.$

The middle panels in \autoref{fig:2DrKE} show the evolution of radial kinetic energy and eccentricity in simulations with eccentricity driven from the outer boundary. Following a transient period, these panels illustrate relaxation to a quasi-steady state maintained by the outer boundary condition. The unfolded polar snapshots of radial velocity shown in the righthand panels of \autoref{fig:2DrKE} illustrate the development of an eccentric distortion in simulation Af25r3D. The colormaps of $u_r$, overlaid by streamlines as in \autoref{fig:A25vr}, show the initial inward propagation of eccentricity (top), the peak in radial kinetic energy achieved at $t\sim2500\Omega_0^{-1}$ (middle), and the final steady-state eccentric disk (bottom). We find that eccentric distortions driven from the outer boundary peak in eccentricity at larger radii than the nonlinear eigenmode solutions to the secular theory. 

\begin{figure*}
    \centering
    \includegraphics[width=\textwidth]{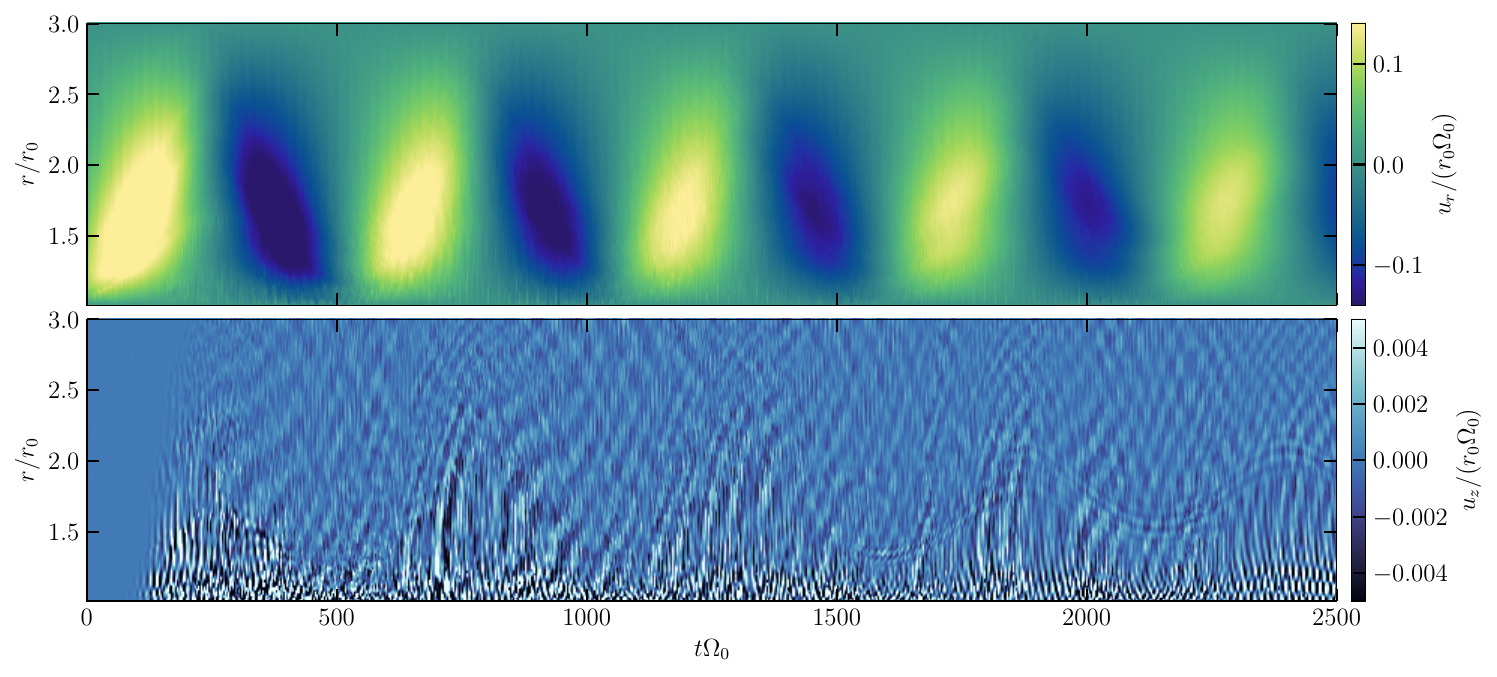}
    \caption{Spacetime diagrams analogous to those shown in \autoref{fig:A25r32D_spct}, but showing mid-plane radial profiles for the 3D simulation A25r3D. The additional spacetime diagram showing midplane vertical velocity (bottom) illustrates the growth of small-scale inertial oscillations close to $r_0$, and their breakdown into subsonic turbulence. 
    }\label{fig:A25r3D_spct}
\end{figure*}

\subsection{3D simulations}\label{sec:3Deig}
Like \autoref{fig:A25r32D_spct}, \autoref{fig:A25r3D_spct} (top) shows radial velocity as a function of radius and time for a simulation, A25r3D, initialized with a free eccentric mode with maximum eccentricity of $0.25$, and run with the semi-rigid vertical boundary conditions described in \autoref{sec:bc}. The eccentric initial condition has simply been copied at each vertical cell (i.e., the vertical profile of eccentricity is initially uniform). In 3D the eccentric disk precesses marginally more slowly than in 2D (compare the final phase of precession between \autoref{fig:A25r32D_spct} and \autoref{fig:A25r3D_spct}, top). This slowdown in precession derives ultimately from the vertical oscillations illustrated by the spacetime diagram of vertical velocity shown in the bottom panel of \autoref{fig:A25r3D_spct}. The nature and consequences of these vertical oscillations provide our main focus. 

\begin{figure*}
    \centering
    \includegraphics[width=\textwidth]{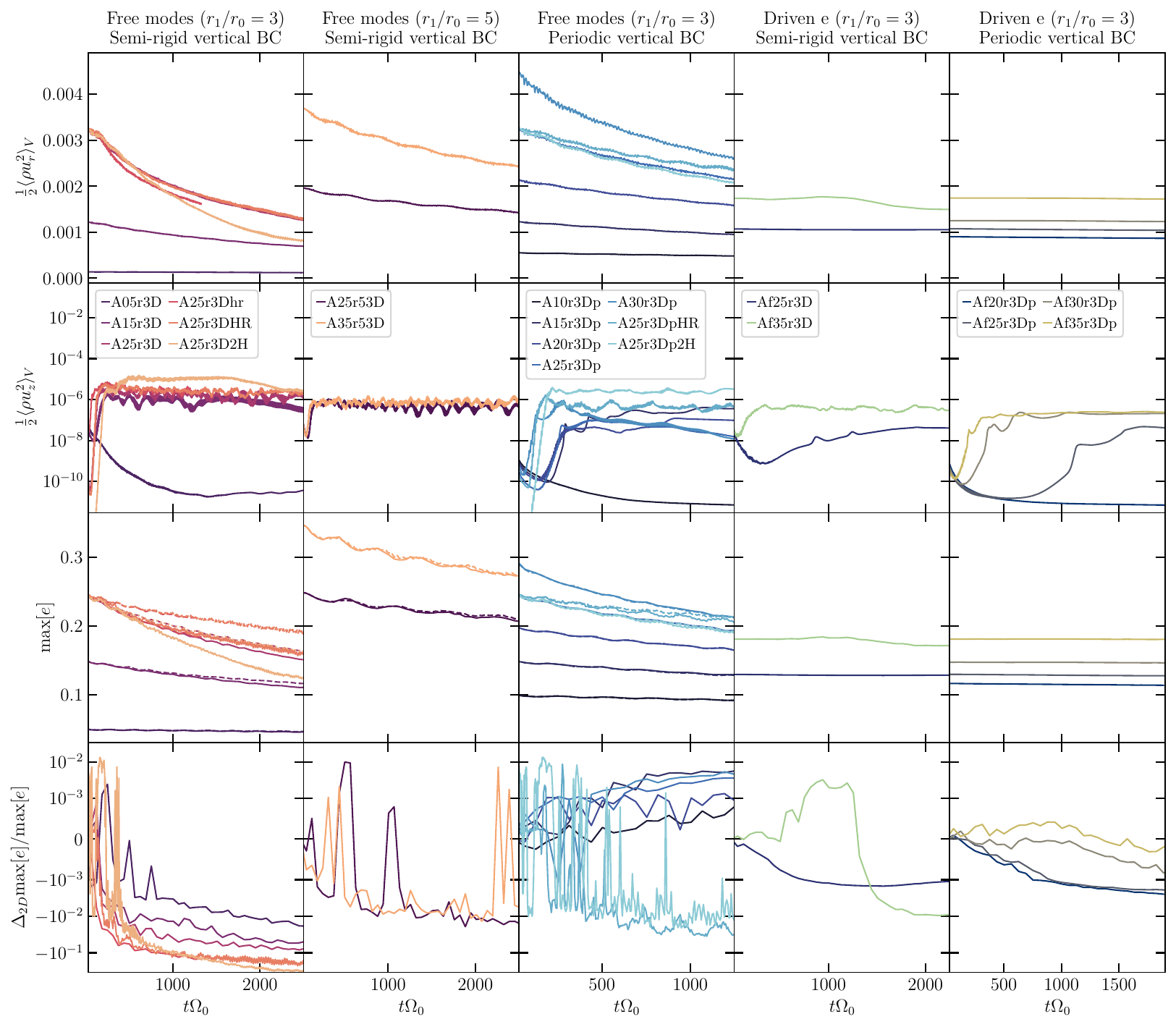}
    \caption{
    Time evolution of integrated quantities for our 3D simulations. Column titles indicate the source of eccentricity (free mode vs. driven distortion), while line colors indicate different eccentricity values or driving amplitudes. From top to bottom, the each row plots the volume-integrated radial kinetic energy, the vertical kinetic energy, the maximum eccentricity, and the (relative) difference in maximum eccentricity between each 3D simulation and its 2D counterpart. The $y$-axes in the final row transition from log to linear scales at $10^{-3}.$ A combination of local and global instabilities intrinsic to the disk distortions drive the growth of vertical oscillations, which in turn modify the background disk eccentricity. 
    }\label{fig:3dsimhist}
\end{figure*}

\begin{figure*}
    \centering
    \includegraphics[width=\textwidth]{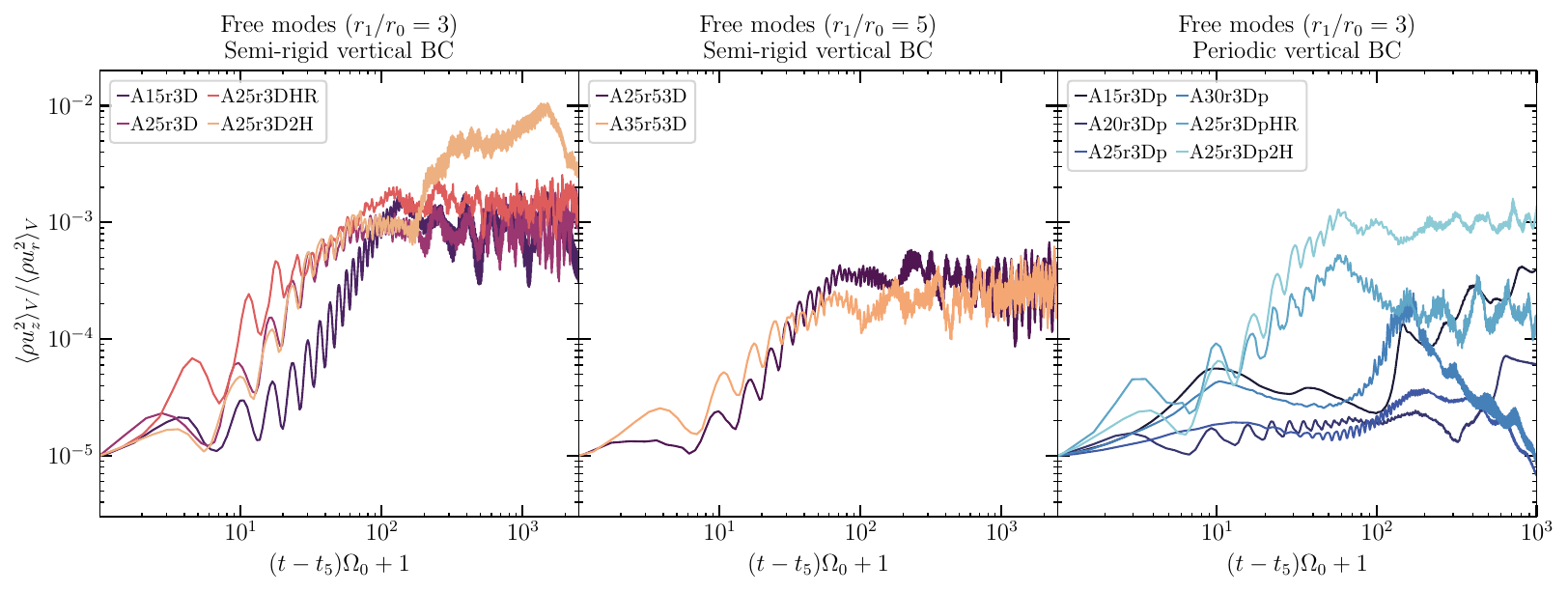}
    \caption{
    Time evolution of the ratio of vertical to radial kinetic energy for freely eccentric simulations showing significant growth in the former. For each simulation, the $x$-axis is shifted by $t_5$, the time during initial growth at which this ratio surpasses $10^{-5}$ (this shift corrects for differences in initial perturbation amplitude for some of the simulations). 
    }\label{fig:keratio}
\end{figure*}

\subsubsection{Eccentricity damping}
\autoref{fig:3dsimhist} summarizes the time evolution of volume-integrated quantities for 3D simulations run with a range of initial conditions, domain sizes, and boundary conditions. 

The panels in the top row of \autoref{fig:3dsimhist} show that our 3D simulations experience a similar decay (with free eccentric modes) or plateau (with driven eccentricity) of radial kinetic energy to the 2D simulations described by \autoref{fig:2DrKE}. Unlike the 2D simulations, the hydrodynamic instability of eccentric disks in three dimensions leads to the additional growth of vertical kinetic energy, shown by the second row of panels in \autoref{fig:3dsimhist}. This growth takes place within a few hundred orbital periods in all but our least eccentric simulations. As shown in \autoref{fig:keratio}, the energy in vertical motions saturates at an amplitude of $\sim 10^{-4}$ to $10^{-3}$ times the energy in radial motions. For most cases this corresponds to root mean squared vertical velocities $(|v_z|^2)^{1/2}\sim 0.02-0.06c_s$ (depending on radial extent and boundary conditions), in rough agreement with the value of $0.03c_s$ found by \citet[][cf. his Section $4.2$]{Papaloizou2005b}. 

Assuming $\rho u_z^2\sim \rho_0\delta u_z^2\propto \exp[2st],$ and fitting an exponential curve to the volume average $\langle\rho u_z^2\rangle_V$ during the initial stage of growth, we find growth rates (listed in Table \ref{tab:3Dsim}) that increase with eccentricity. These growth rates are comparable to, but smaller than the maximum values of local growth rate indicated by \autoref{eq:smax} (these maximum values are always achieved at the inner boundary, due to the influence of the eccentricity gradient).

The character of the saturation following exponential growth depends on the vertical boundary condition, the vertical extent of the simulation, and the source of eccentricity. In simulations with semi-rigid vertical boundaries at $\pm H_0=\pm c_s/\Omega(r_0)$ (\autoref{fig:keratio} left, middle), the ratio of vertical to radial kinetic energy fluctuates around a roughly static value over short timescales. Our simulations with periodic vertical boundary conditions (\autoref{fig:keratio}, right) and/or a larger vertical extent (lightest curves in the left and right panels of \autoref{fig:keratio}) show larger variations in this ratio over longer timescales. We explore the dynamics underlying these differences in saturation in the following subsections. 

\begin{figure*}
    \centering
    \includegraphics[width=\columnwidth]{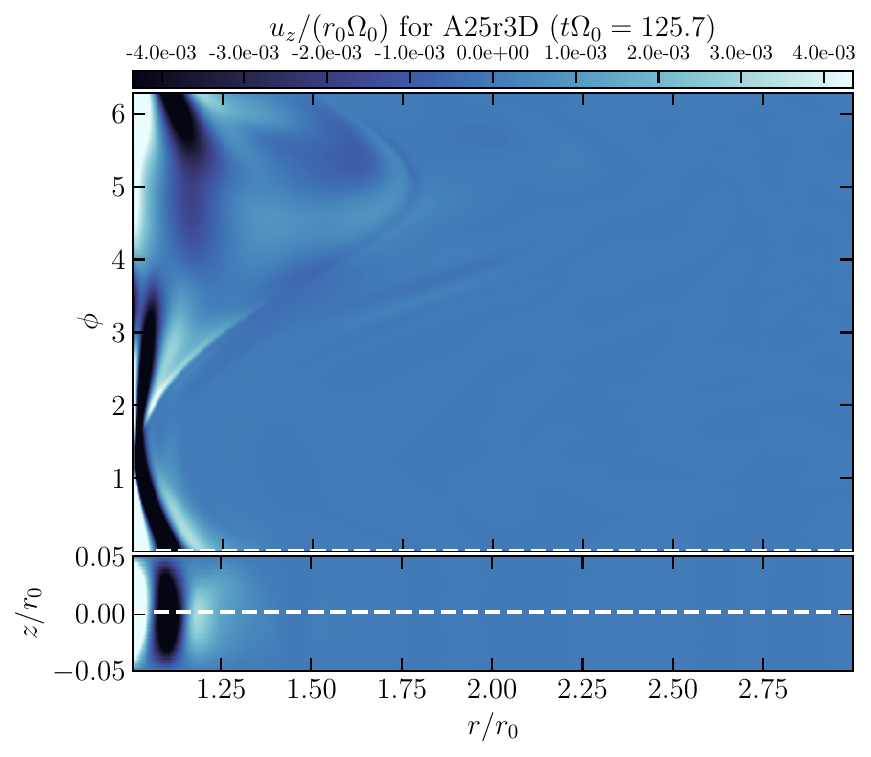}
    \includegraphics[width=\columnwidth]{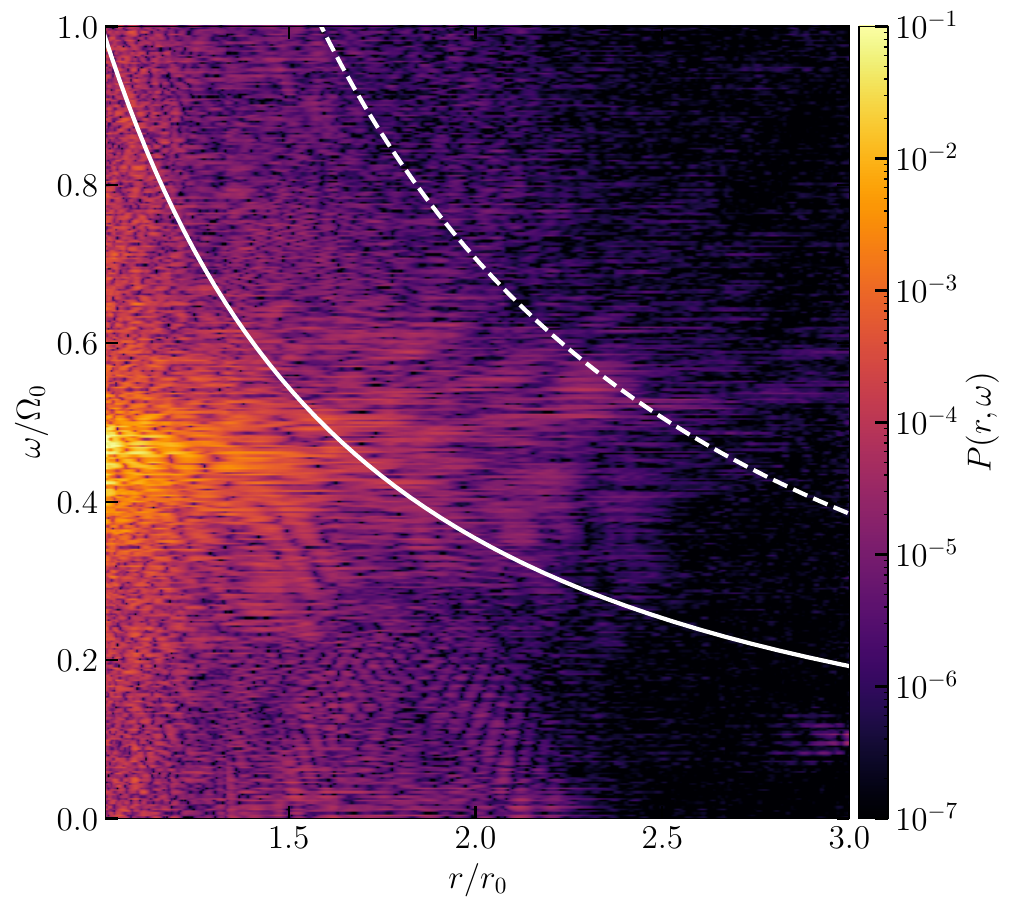}
    \caption{Left: Snapshot of simulation A25r3D showing unfolded polar and vertical slices of vertical velocity during the initial stages of growth. The dashed line in each panel indicates the $\phi/z$ slicing location of the other. Right: PSD $P(r,\omega)$ computed from the temporal Fourier transforms of vertical velocity profiles $u_z(r,\phi=0,z=0)$, sampled at a cadence of ten times every orbital period from $t\Omega_0=100$ to $2500$. Solid and dashed white lines indicate profiles of $\Omega(r)$ and $2\Omega$  (respectively) for a circular disk.}\label{fig:A25r3D_snap}
\end{figure*}

Regardless of the nature of the disks' vertical oscillations, their growth comes at the expense of the free energy stored in the disk's eccentricity: the panels in the third row of \autoref{fig:3dsimhist} plot maximum eccentricity for both our 3D simulations (solid lines), and their corresponding 2D analogue simulations (dashed lines). The final row then plots the relative difference between these eccentricity measurements (computed at identical times), with negative values corresponding to faster eccentricity decay in 3D than in 2D. In all but a subset of simulations run with periodic vertical boundary conditions, the instabilities of our eccentric disks lead to an enhancement in their own damping. 

\begin{figure*}
    \centering
    \includegraphics[width=\textwidth]{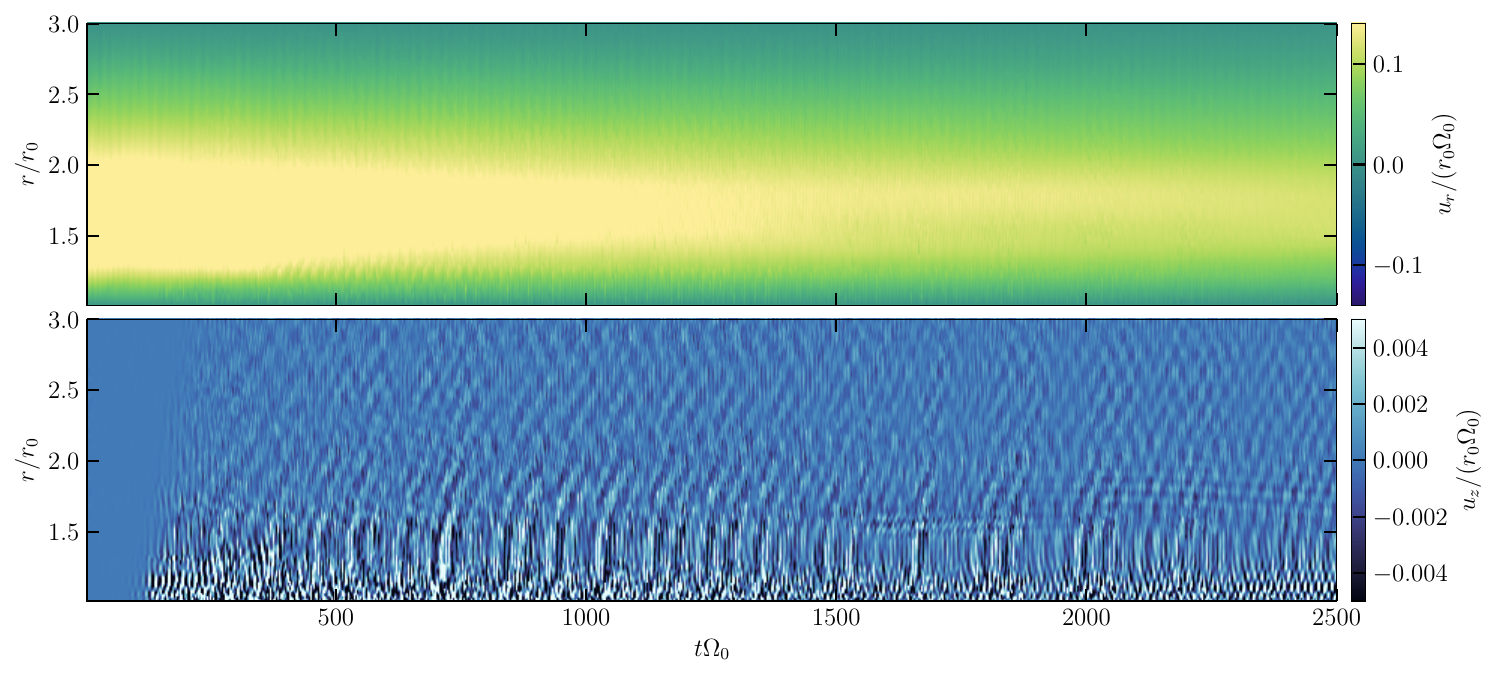}
    \caption{Similar spacetime diagrams to those shown in \autoref{fig:A25r3D_spct}, but with mid-plane radial (top) and vertical (bottom) velocities evaluated at an azimuthal angle $\phi$ chosen instantaneously to coincide the $\phi$ of the maximum radial velocity in the simulation domain (this $\phi$ precesses with the eccentric disk). 
    }\label{fig:A25r3D_pspct}
\end{figure*}

\subsubsection{Inertial wave excitation}
Parametrically driven inertial waves produce the rapid fluctuations in vertical velocity shown in the bottom panel of \autoref{fig:A25r3D_spct}. This spacetime diagram illustrates wave propagation throughout the extent of the disk, as well as standing waves, but with amplitudes largest near the inner radial boundary. \autoref{fig:A25r3D_snap} unpacks some of the spatial and temporal characteristics of the inertial waves from A25r3D. The lefthand snapshot shows $r-\phi$ and $r-z$ slices of vertical velocity during the initial growth of the parametric instability. The vertical slice illustrates vertical wavelengths $\lambda_z\sim 4H_0=0.2r_0.$ We then observe radial wavelengths of $\lambda_r\sim2.5H_0\sim0.1r_0$, which compare well with the expectation that the maximal growth rate given by \autoref{eq:smax} occurs for $(k_rH_0)^2\sim 3[(k_zH_0)^2 - 1/4].$ 

The righthand panel in \autoref{fig:A25r3D_snap} shows power spectral density computed from mid-plane radial profiles of vertical velocity, measured in the inertial frame (normalized by the maximum value of the PSD). The solid white line plots the orbital frequency of a circular, Keplerian disk, while the dashed line plots twice this frequency. The PSD for this simulation shows peaks at roughly half the local orbital frequency close to both the inner and outer boundaries. These frequencies and localizations both agree with local analyses (i.e., \autoref{eq:smax}), which predict the largest growth rates at $r_0$ and $r_1$ (see \autoref{fig:A25vr}, bottom). The appearance of a peak in power at marginally less than half the local orbital frequency near the inner boundary (i.e., at $\sim0.4-0.45\Omega_0$) indicates that the dominant growth is associated with inertial waves localized to marginally larger radii (where the local orbital frequency is smaller than $0.5\Omega_0$).

Turbulence develops at a few tens of orbital periods in all of our simulations, and introduces power at a variety of other frequencies and radii. However, the fastest growing inertial waves remain apparent in nearly all cases. \autoref{fig:A25r3D_pspct} shows similar spacetime diagrams to \autoref{fig:A25r3D_spct}, but with radial (top) and vertical (bottom) velocities evaluated at azimuthal angles chosen to precess with the background eccentricity. In this precessing frame the decay in radial velocity amplitude clearly illustrates the decay of eccentricity. Meanwhile, diagonal features at $r/r_0\gtrsim1.5$ in the vertical velocity spacetime diagram indicate inward and outward wave propagation. Interior to $r/r_0\lesssim1.5,$ nearly vertical patterns in the $u_z$ spacetime diagram additionally indicate the formation of standing waves. The checkerboard patterns close to the outer boundary correspond to the lower frequency inertial waves indicated by the weak band of power at $\omega\sim0.5\Omega(r_1)$ shown in \autoref{fig:A25r3D_snap} (right).

\begin{figure*}
    \centering
    \includegraphics[width=\columnwidth]{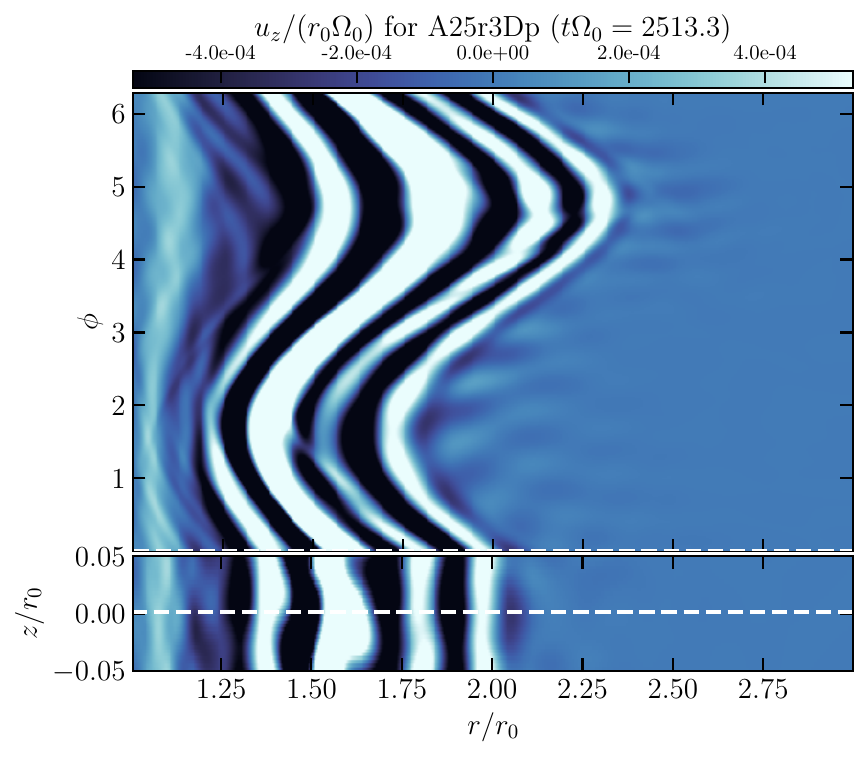}
    \includegraphics[width=\columnwidth]{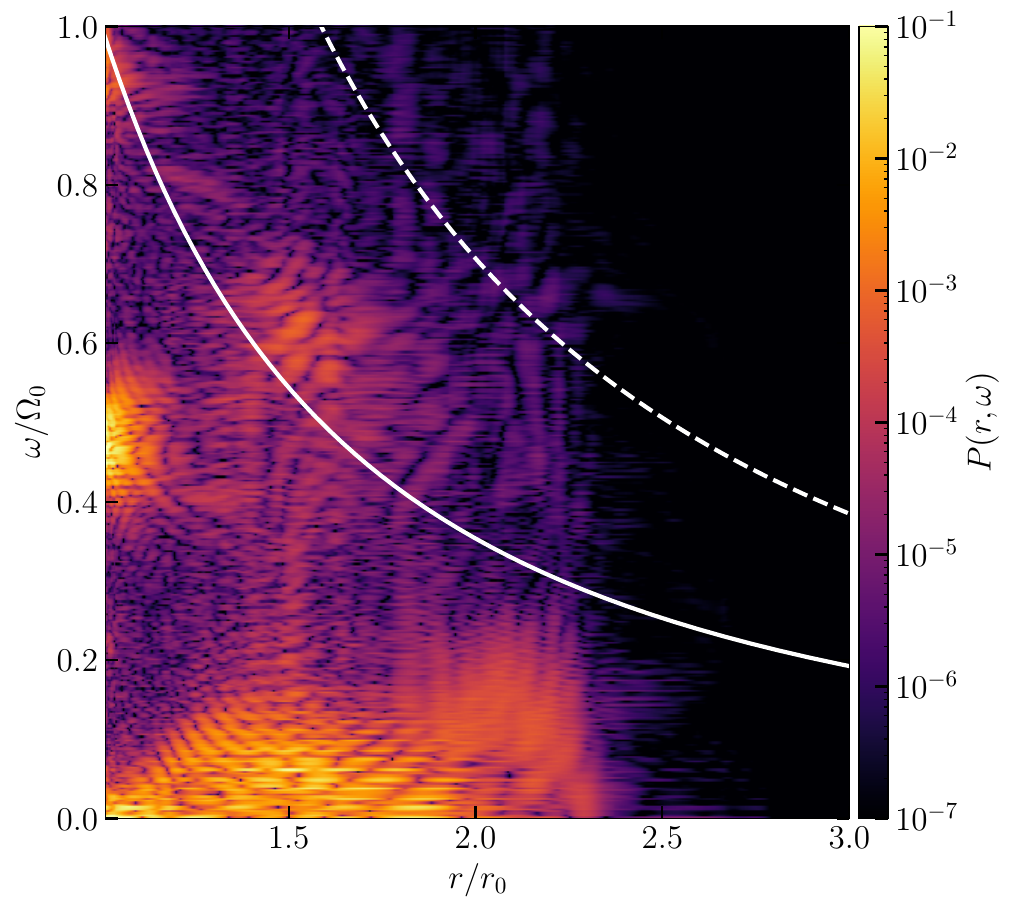}
    \caption{Left: $r-\phi$ and $r-z$ snapshots of $u_z$ like those shown in \autoref{fig:A25r3D_snap}, taken at the end of run-time for A25r3Dp, an simulation run with periodic (rather than rigid) vertical boundary conditions. Right: power spectral density $P(r,\omega)$ calculated from $u_z(r,\phi=0,z=0,t)$ spacetime data from the same simulation.}\label{fig:A25r3Dp_snap}
\end{figure*}

\subsubsection{Periodic vertical boundary conditions}\label{sec:elev}
Our simulations run with periodic vertical boundary conditions (the natural choice for a cylindrical, unstratified model) show similar inertial wave excitation to those run with semi-rigid vertical boundary conditions. However, in all but the vertically extended A25r3Dp2H, their evolution at later times is dominated by the development of elevator flows like the one shown in \autoref{fig:A25r3Dp_snap} (left). These elevators persist as steady flows that are time-invariant in the frame precessing with the background eccentric disk. Tracing the most eccentric streamlines in the disk, they comprise multiple azimuthal Fourier modes. The doppler shifting of multiple azimuthal wavenumbers to an inertial frame then produces patterns in power spectral density like those shown at frequencies $\omega\lesssim0.2\Omega_0$ in \autoref{fig:A25r3Dp_snap} (right). 

In simulations with periodic vertical boundary conditions imposed at $z=\pm H_0$, the formation of elevator flows generally coincides with the saturation of vertical kinetic energy growth (at lower values than in simulations with rigid vertical boundary conditions; see the rightmost panel of \autoref{fig:keratio}). This suggests that they act similarly to the zonal flows identified by \citet{Wienkers2018}, throwing the small-scale inertial oscillations out of resonance before wave-breaking can occur. 

\subsubsection{Vertically extended domains}
\begin{figure*}
    \centering
    \includegraphics[width=\textwidth]{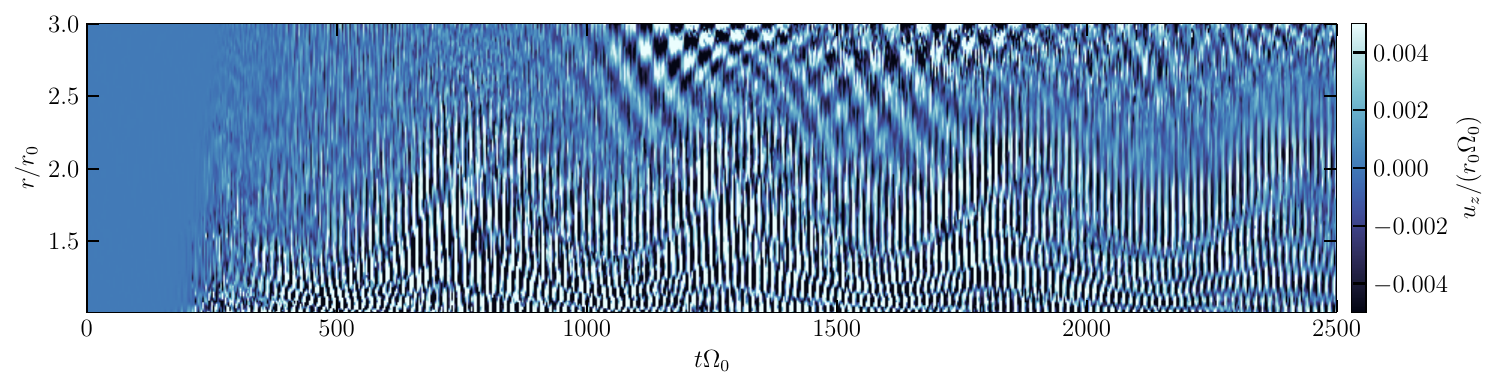}
    \caption{Same as \autoref{fig:A25r32D_spct} (bottom), but for simulation A25r3D2H (which has twice the vertical extent).}\label{fig:A25r3D2H_uzspct}
\end{figure*}

Our simulations with a larger vertical extent exhibit larger values of saturated vertical kinetic energy, and stronger peaks in analogous PSDs to the one shown in \autoref{fig:A25r3D_snap}. \autoref{fig:A25r3D2H_uzspct} shows a spacetime diagram of vertical velocity in A25r3D2H, a simulation with twice the vertical extent of the fiducial run A25r3D. Stronger low-frequency oscillations in the outer disk are apparent even in the time domain. \autoref{fig:A25r3D2H_uzspct} also exhibits more radially extended standing wave features.

These differences may arise because a larger vertical extent permits growing modes with longer vertical and radial wavelengths, which are, in turn, less affected by grid diffusion at a fixed resolution. Although the dominant, fastest growing inertial waves initially share the same vertical wavelengths in simulations with different vertical scales, we do observe modes with longer wavelengths (in particular, $k_zH_0\sim\pi/4$ rather than $\pi/2$) at later times in the vertically extended simulations. The larger vertical extent also reduces the impact of the elevator flows described in the previous section; they grow in A25r3Dp2H, but do not lead to the same reduction in saturated vertical kinetic energy (see 
\autoref{fig:keratio}, right). 

Our taller simulations do not necessarily provide a more accurate picture of the parametric instability in real accretion disks. The cylindrical model we adopt is an approximation that is best suited to vertical domains significantly less than $H$, not to those that extend to $\pm 2H$. Moreover, the radial propagation behavior of inertial waves differs between vertically stratified and unstratified models. The `corrugation' mode, in particular, is inadequately described in the latter model, which treats its Lindblad resonance as a true turning point and thus artificially favors standing-mode formation. The discrepancies between our simulations with different vertical extent highlight the limitations of our numerical framework, and motivate future investigation of vertically stratified eccentric disk models.

\begin{figure*}
    \includegraphics[width=\textwidth]{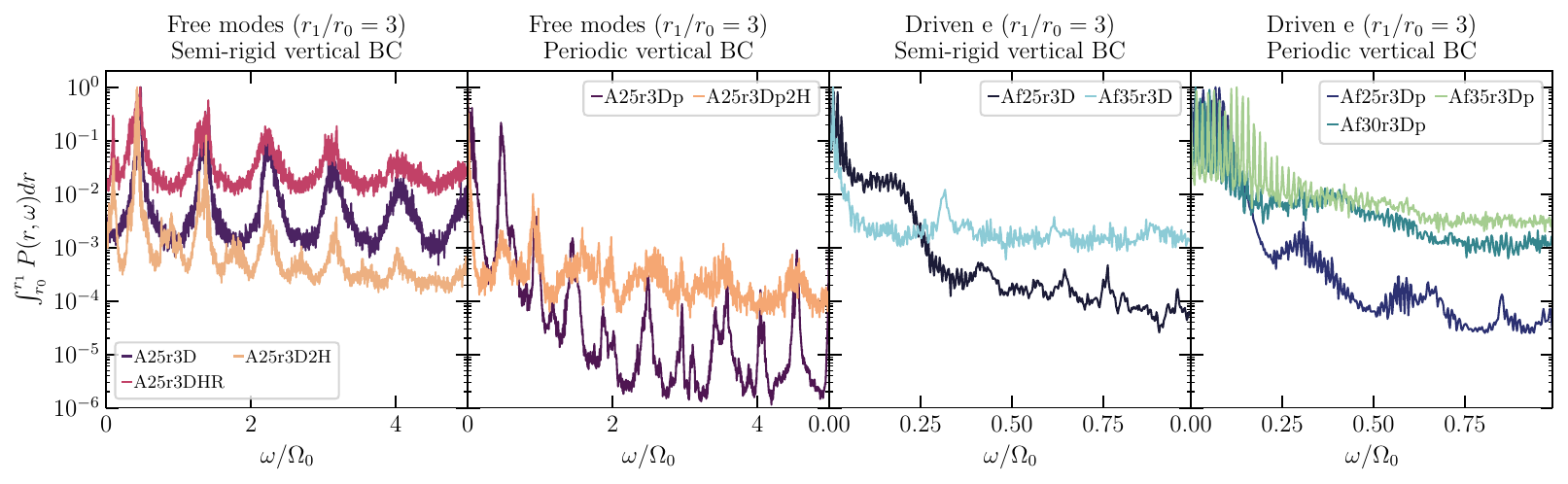}
    \caption{
    Power spectral densities, integrated over radius for the most vertically oscillatory 3D simulations. The left two panels are computed from data sampled as described in \autoref{fig:A25r3D_snap}, while the right two panels were computed from spacetime data sampled twice an orbital period. Each integrated PSD is normalized by its maximum value. The parametric instability produces peaks in power at harmonics of $\sim0.5$ times the local orbital frequency (dominated by waves excited in the inner disk). Meanwhile, global instability produces power at lower frequencies. 
    }\label{fig:allsim_1DuzPSD}
\end{figure*}

\subsubsection{Impact of resolution}
Although our initial conditions produce a cleaner manifestation of the parametric instability than found in the simulations of \citet{Papaloizou2005b}, we do not claim to resolve it completely. As shown in the leftmost panels of \autoref{fig:3dsimhist} and \autoref{fig:keratio}, A25r3DHR and A25r3DpHR (simulations with the resolution doubled in all dimensions) exhibit enhancements to the saturated vertical kinetic energy and eccentricity damping achieved in A25r3D and A25r3Dp (respectively). Additionally, although A25r3DpHR develops nearly identical elevator flows to A25r3Dp, this higher resolution run simultaneously sustains larger amplitude turbulent fluctuations at smaller spatial scales (in turn leading to the sustained vertical kinetic energy shown in the right panel of \autoref{fig:keratio}).

We attribute these enhancements to the growth of inertial waves with larger wavenumbers (shorter wavelengths), which are permitted by the dependence of parametric instability growth rates on the ratio of radial to vertical wavenumber, rather than the value of either. Fully resolving the parametric instability in a global simulation would likely require extremely high resolution \citep{Cui2022}. We see very little difference between the properties of the \emph{longest} wavelength features in A25r3D and A25r3DHR, or in A25r3Dp and A25r3DpHR. However, the possible excitation of shorter and shorter wavelength inertial waves with increasing resolution implies our simulations provide only a lower bound on the virulence of the parametric instability.

\subsubsection{Slow global instability}

\autoref{fig:allsim_1DuzPSD} summarizes the variability in a representative selection of 3D simulations. As a function of frequency, each panel plots power spectral density $P(r,\omega)$ integrated over the radial domain, with each integral normalized by its maximum value. 

Regardless of vertical boundary condition, the simulations initialized with a free eccentric mode show peaks in power at (harmonics of) roughly half the inner orbital frequency that clearly identify parametric inertial wave excitation. Additionally, the panels in \autoref{fig:allsim_1DuzPSD} all show enhanced power at low frequencies $\lesssim0.1\Omega_0$. This is even more apparent in our simulations with eccentricity driven from the outer boundary (two rightmost panels in \autoref{fig:allsim_1DuzPSD}), in which low-frequency power dominates. This low-frequency power is due partially to the elevator flows discussed in \autoref{sec:elev}. However, it also derives from another family of global, vertically structured flows that grow in all of our simulations.

\begin{figure*}[ht!]
    \centering
    \includegraphics[width=\textwidth]{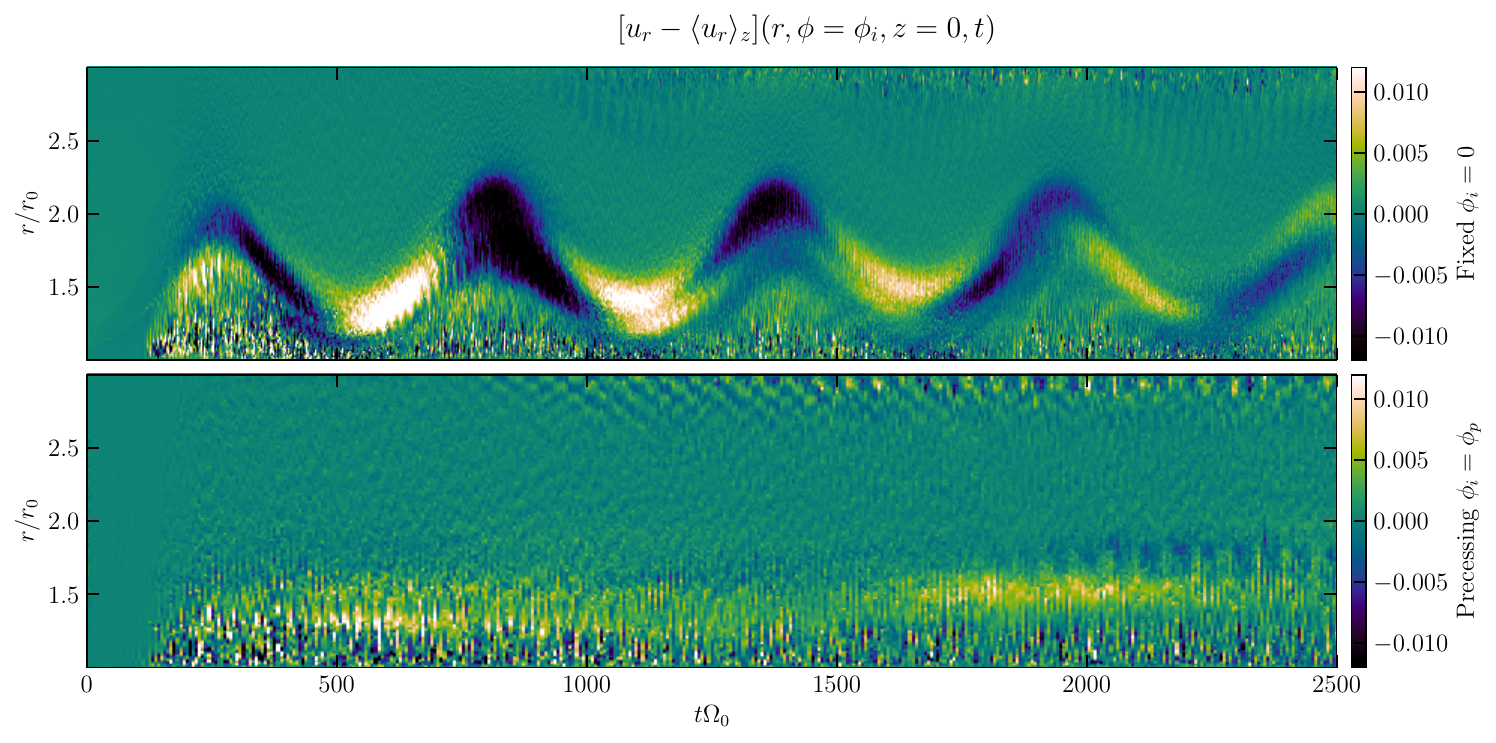}
    \caption{Spacetime diagrams from A25r3DHR, showing mid-plane radial velocity fluctuations $u_r-\langle u_r\rangle_z$ from the background eccentric disk. The top diagram shows radial profiles taken at a fixed azimuthal angle, while for the bottom diagram $\phi$ has been chosen to precess with the eccentric disk. The spacetime diagrams show the evolution of a global mode with a very low frequency in the precessing frame.}\label{fig:A25r3D_dzvgUr_spct}
\end{figure*}

The spacetime diagrams in \autoref{fig:A25r3D_dzvgUr_spct} show radial deviations from the vertically averaged radial velocity in A25r3DHR. Subtracting this vertical average removes the eccentric initial condition. The top panel shows a spacetime diagram constructed from mid-plane radial profiles measured in the inertial frame, while the bottom panel shows spacetime evolution in the same precessing frame as \autoref{fig:A25r3D_pspct}. \autoref{fig:A25r3D_dzvgUr_spct} demonstrates the oscillation of a global mode with a very low frequency in the precessing frame. 

\begin{figure*}[ht!]
    \centering
    \includegraphics[width=\textwidth]{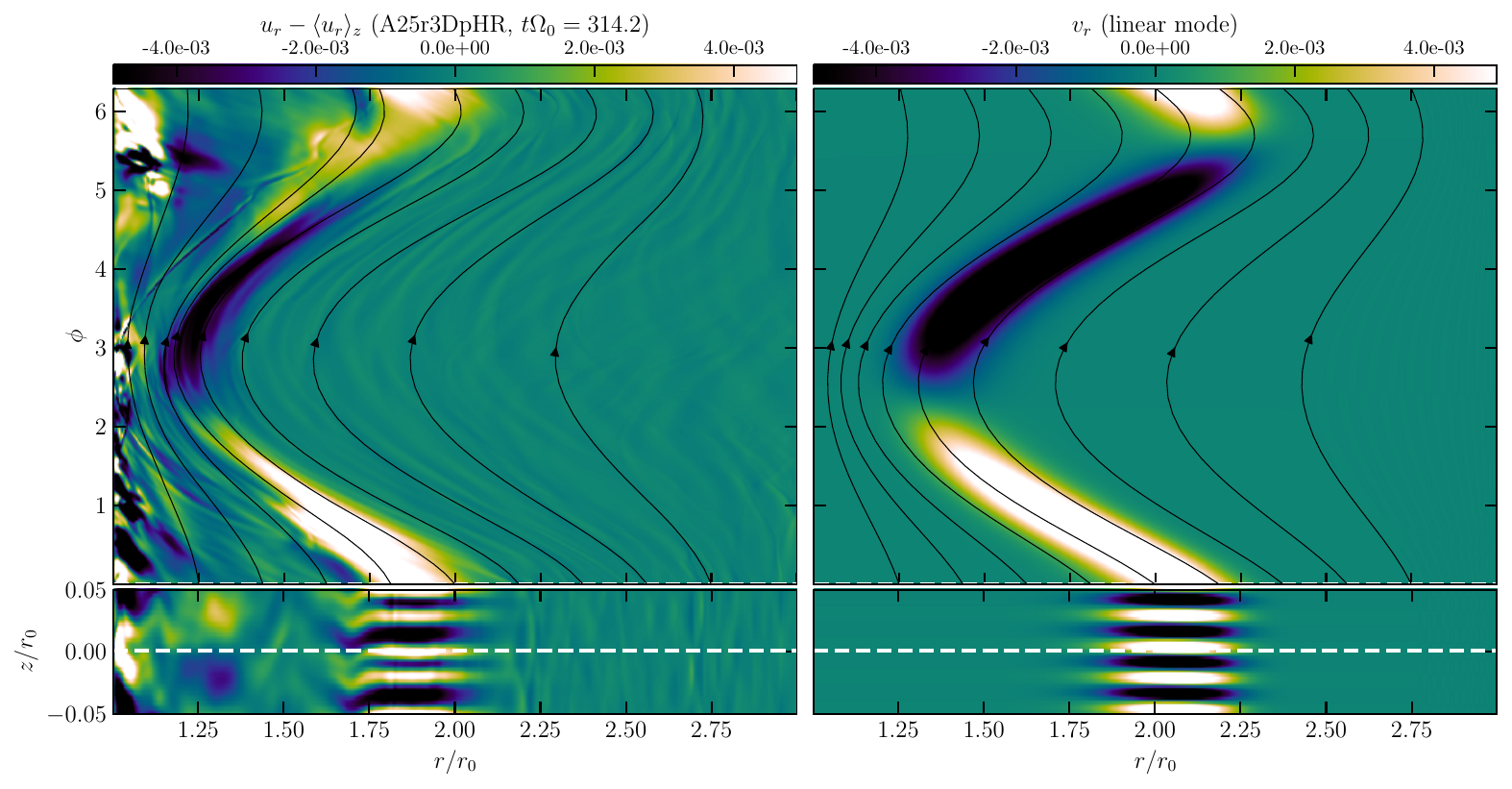}
    \caption{Left: snapshots of the deviation $u_r-\langle u_r\rangle_z$ in A25r3DpHR, taken toward the beginning of the simulation runtime (after $\sim50$ orbital periods). Right: analogous plot for a linear mode computed with $k_z=4\pi/H$ for a secular eccentric disk solution with the same instantaneous $\max[e]\approx0.23$ (but not exactly the same eccentricity profile) as the simulation on the left.}
    \label{fig:A25r3Dp2H_globmode}
\end{figure*}

\autoref{fig:A25r3Dp2H_globmode} (left) shows a snapshot of radial velocity perturbation in A25r3DpHR, taken shortly after the saturation of vertical kinetic energy growth. The vertical and midplane slices illustrate a mode with $k_z\sim4\pi/H_0$ in the vertical direction, and the equivalent of an $m\sim1$ structure from the perspective of the eccentric streamlines (i.e., in orbital coordinates). \autoref{fig:A25r3Dp2H_globmode} (right) shows a linear normal mode computed (as described in Appendix \ref{app:lin}) for an eccentric disk solution with the same maximum eccentricity as instantaneously measured for the snapshot on the left. Moving backward from eccentric to circular background states, we in turn identify such linear modes with secondary eccentric modes involving a vertical variation in eccentricity (i.e., introducing the $z-$dependence $e\propto\exp[\text{i}k_zz]$).

\begin{figure*}[ht!]
    \centering
    \includegraphics[width=\textwidth]{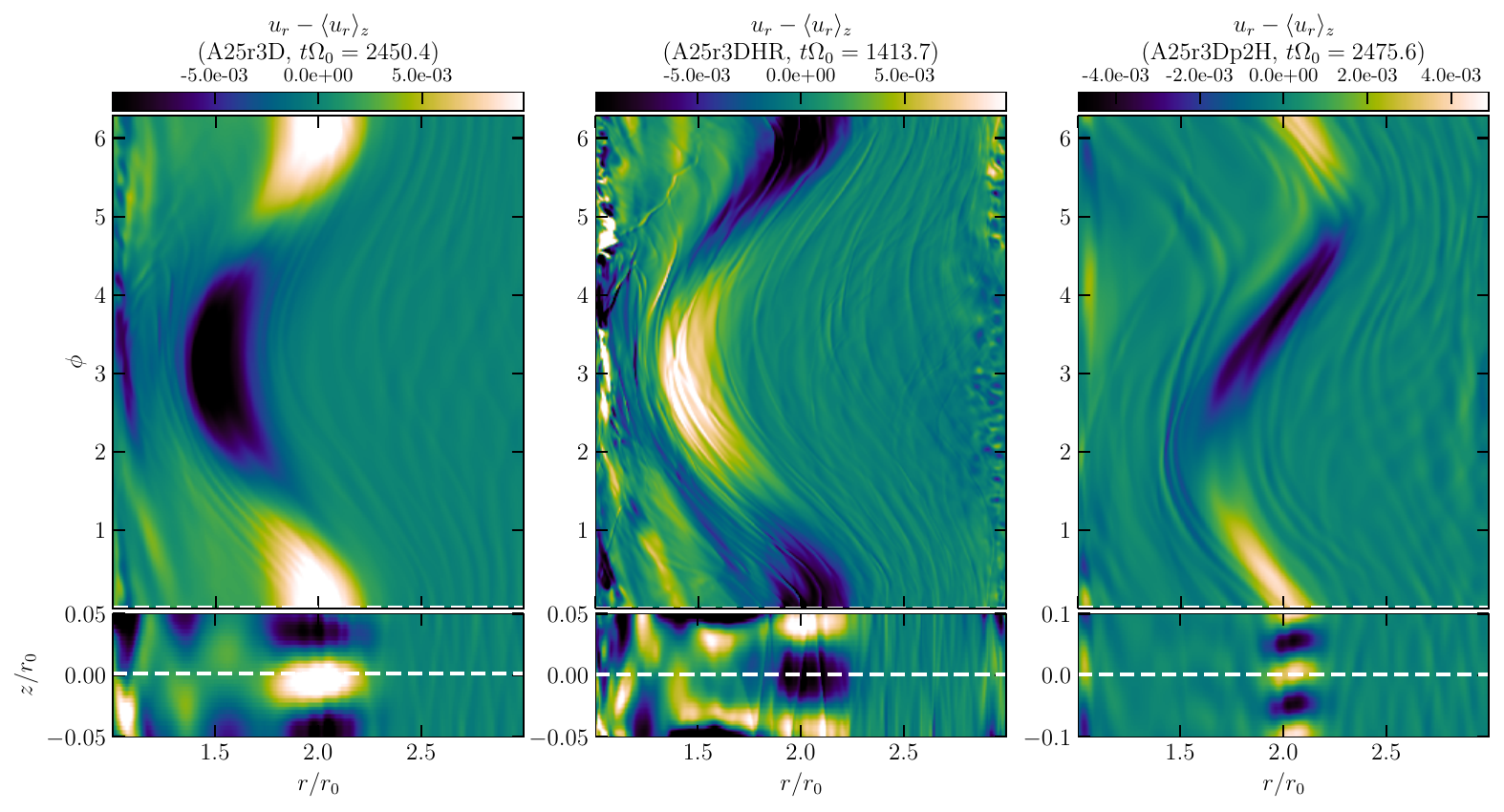}
    \caption{Snapshots illustrating the global modes that develop at late times in A25r3D (left), A25r3DHR (middle), and A25r3Dp2H (right). These exhibit longer vertical wavelengths than the initially excited mode shown in \autoref{fig:A25r3Dp2H_globmode} (left).}
    \label{fig:mgal}
\end{figure*}

\begin{figure*}[ht!]
    \centering
    \includegraphics[width=\textwidth]{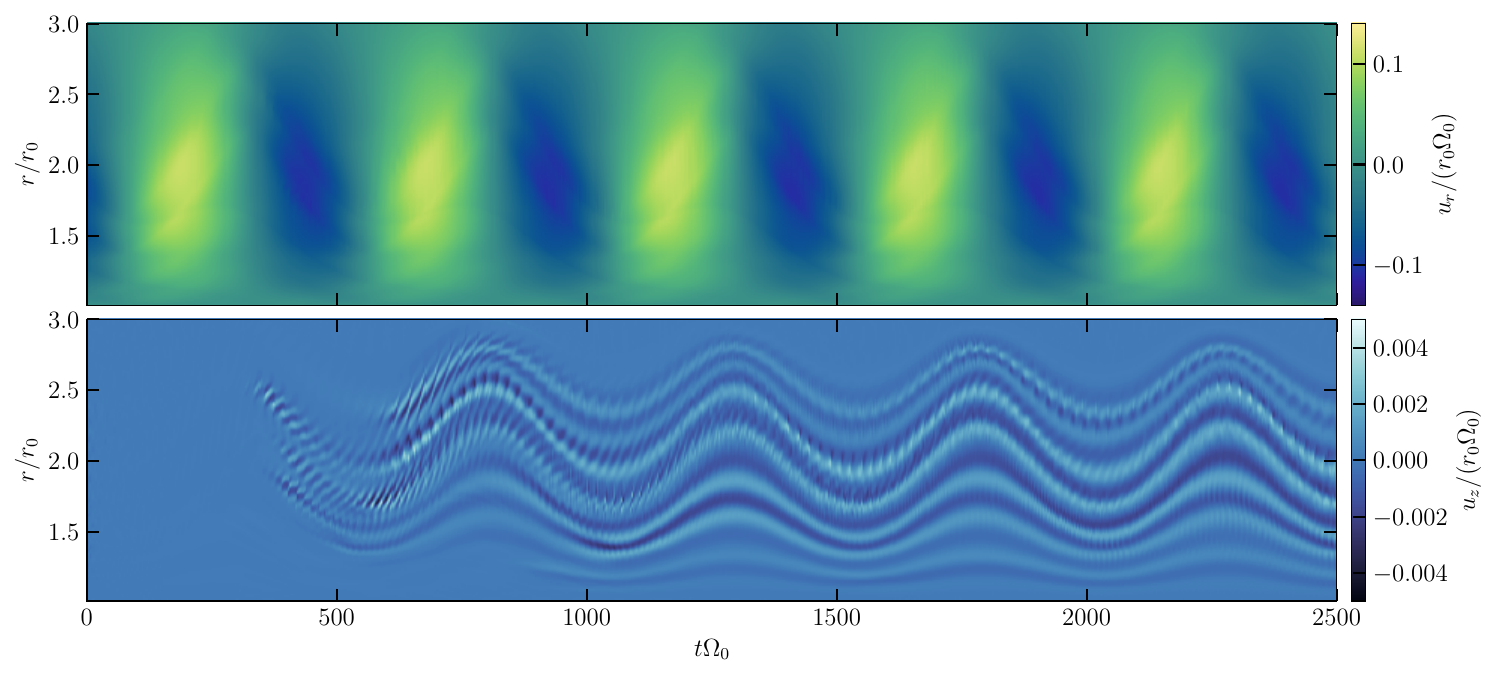}
    \caption{Same as \autoref{fig:A25r3D_spct}, but for a simulation with driven eccentricity (outer BC set by an eccentric mode with $\max[e]=0.3,$ $r_1/r_0=3$), and periodic vertical boundary conditions. The growth in vertical kinetic energy is dominated in this case by globally unstable eccentric modes, which saturate rapidly in elevator flows, and are beset by shear instabilities.}\label{fig:Af35r3Dp_UzSpct}
\end{figure*}

Linear normal modes like the one shown in \autoref{fig:A25r3Dp2H_globmode} (right) closely resemble the global modes that manifest in our simulations in a number of respects: they have similarly low frequencies in the precessing frame ($\omega\sim0.025\Omega_0$ for the case shown in \autoref{fig:A25r3Dp2H_globmode}, right), an identical phase relationship between radial and azimuthal velocity components, and initially similar spatial structure.

However, rigid vertical boundary conditions appear to modify the global modes' properties, reducing their (already low) frequencies, confining them closer to the inner boundary, and lengthening their vertical wavelengths. At late times the global modes in simulations with periodic vertical boundary conditions also develop longer vertical wavelengths (with vertical wavenumbers $k_z\sim\pi/H_0$), while maintaining the same horizontal structure (see \autoref{fig:mgal}). In contrast, the linear modes' horizontal wavelengths increase with their vertical wavelengths (see Figure \ref{fig:kzvar_n1}).

In simulations with periodic vertical boundary conditions, the global modes are low frequency but transient features. After saturating at similar amplitudes to those indicated by \autoref{fig:A25r3D_dzvgUr_spct}, they give way to elevator flows. Their repeated excitation then contributes to more and more extensive elevator flow development. 

In simulations with driven eccentricity, low growth rates for the inertial wave parametric instability leads to variability entirely dominated by global mode growth and elevator flow development (see \autoref{fig:allsim_1DuzPSD}, right). Rather than parametric instability, in simulations with driven eccentricity we find that the unstable global modes become unstable to secondary shear instabilities. These parasites can be seen in the bottom panel of \autoref{fig:Af35r3Dp_UzSpct}, which shows spacetime diagrams of mid-plane radial (top) and vertical (bottom) velocity (sliced in the inertial frame) in the simulation Af30r3Dp.

\subsection{Vertical eccentricity variation}\label{sec:disk}
We have not identified an unambiguous source of instability for the global, vertically structured eccentric modes that grow ubiquitously in our simulations. Eigenmode calculations like the one shown in \autoref{fig:A25r3Dp2H_globmode} (right) produce purely (albeit very slowly) oscillating modes with vanishing growth rates, at least when the freely eccentric disks that serve as our initial conditions are used as background mean flows. One possibility is that the secondary eccentric modes grow due to nonlinear interactions associated with the saturation of the parametric instability. Another alternative is that the global modes are destabilized by the development of a small twist; although the eccentric distortions in our simulations persist as coherent flows, they are not perfect standing modes. We find that arguments of pericentre (computed for each grid cell and binned by semimajor axis) show variations of $\gtrsim0.01-0.1$ radians from the mean value throughout the simulation domain. 

\begin{figure}
    \centering
    \includegraphics[width=\columnwidth]{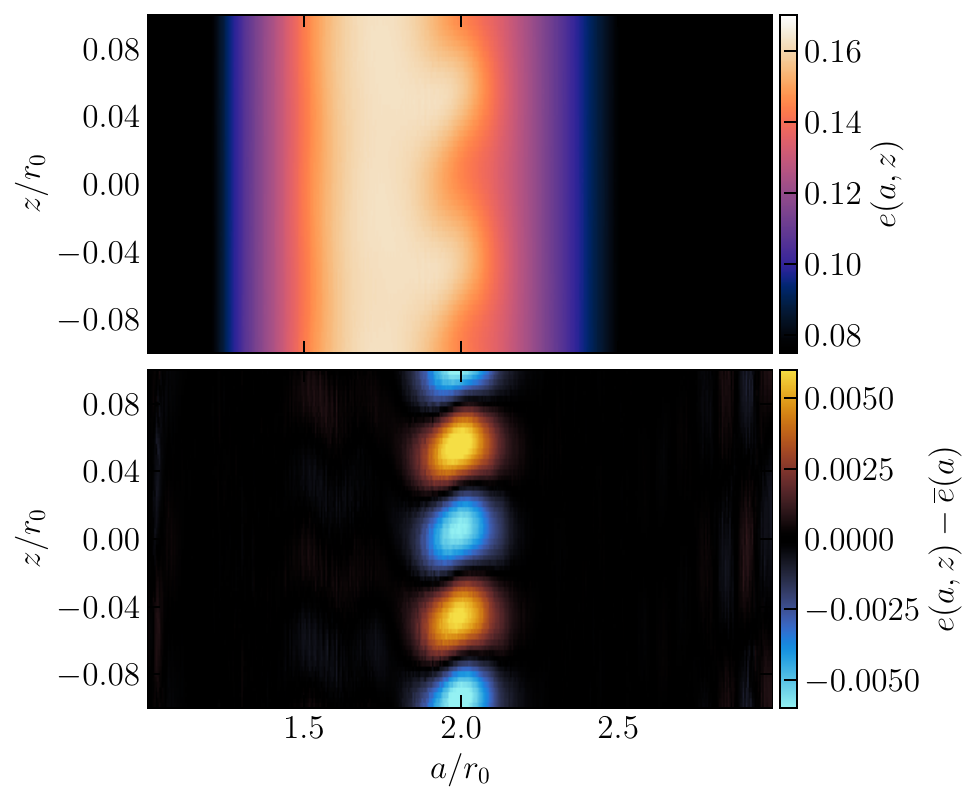}
    \caption{
    Top: eccentricity as a function of semimajor axis and cylindrical $z$, computed for A25r3Dp2H at $t\Omega_0\simeq2470$. Bottom: as in the top panel, but with the background eccentricity profile subtracted. 
    }\label{fig:A25r3Dp2H_erz}
\end{figure}

Regardless of their origin, the secondary distortions alter the (initially uniform) vertical profile of eccentricity. The top panel of \autoref{fig:A25r3Dp2H_erz} shows a snapshot of eccentricity as a function of semimajor axis and height in A25r3Dp2H, computed at the same time as the snapshot shown in \autoref{fig:A25r3Dp2H_globmode}, left. Here we have binned eccentricities by their semimajor axes for all $r$ and $\phi$ but not $z$. Near its maximum at $a/r_0\simeq2$, the eccentricity shows a kinked vertical profile, with $e$ varying noticeably as a function of $z$. The bottom panel of \autoref{fig:A25r3Dp2H_erz} demonstrates this more explicitly, showing $e(a,z)$ minus the profile of eccentricity binned by semimajor axis for all $r,$ $\phi$, and $z.$ This eccentricity perturbation shows a clear modal structure in $z$.

Extrapolating from our vertically local simulations to eccentric disks with vertical gravity is possibly perilous, but interesting. The growth of $k_z\not=0$ eccentric modes in our simulations could correspond to an adjustment of initially uniform vertical eccentricity profiles in full 3D, and/or a warping of initially flat eccentric disks. Such warping might connect with the excitation of bending waves revealed in local models of eccentric disks by \citet{Held2024}. Alternatively, the presence of vertical oscillatory flows associated with scale height variation along eccentric streamlines---together with the enhanced parametric instability that these flows drive---may suppress the growth of secondary distortions. However, the ubiquity of the low-frequency global modes in all of our simulations suggests that they may be robust, if subtle, features of eccentric disk evolution.  

\section{Summary and Conclusions}\label{sec:conc}
This paper characterizes the growth and nonlinear saturation of hydrodynamic instabilities in radially global simulations of strongly eccentric accretion disks. We build on previous investigations by initializing our simulations with fully nonlinear eccentric disk equilibria, which in turn provide a clearer view of eccentricity-driven dynamics. 

The eccentric distortions in our simulations initially excite inertial waves via a parametric instability. We find that this instability saturates when the kinetic energy in vertical motions is roughly $\sim10^{-3}$ times the kinetic energy in radial motions, providing a weak but non-negligible damping for eccentricity. We find growth rates, frequencies, and regions of localization that agree with the predictions of local theory for the eccentric disk parametric instability. However, nonlinear saturation involves a combination of wave propagation and standing mode formation that is intrinsically global. Explorations of parametric instability in simulations of eccentric disks that include vertical gravity provide an important avenue for future work. 

Our simulations also exhibit the growth of lower frequency, global modes that we identify (through $2+1$D eigenmode calculations) as secondary eccentric distortions with varying vertical profiles of eccentricity (i.e., nonzero $k_z$ in our unstratified model). In simulations with periodic vertical boundary conditions, these global modes are accompanied by elevator flows involving purely upward or downward motion, which in turn suppress the parametric instability. In simulations in which we suppress these elevator flows with a rigid vertical boundary condition, or reduce their influence by increasing the vertical extent of the simulation domain, the secondary eccentric modes persist alongside the inertial wave turbulence. Regardless of boundary conditions, they rearrange the disks' structure, producing profiles of eccentricity that vary in the vertical direction (see \autoref{fig:A25r3Dp2H_erz}). We hypothesize that analogous modes in vertically global (i.e., density stratified) models may act to rearrange the vertical structure of initially flat eccentric disks.  

\section*{Acknowledgements}
We thank the anonymous reviewer, who provided helpful comments that improved the quality of the paper. JWD acknowledges financial and computational support provided by the Cambridge Philosophical Society and the Canadian Institute for Theoretical Astrophysics (CITA) during the initial development of this paper. HNL and GIO are supported by the Science and Technology Facilities Council (STFC) through grant ST/X001113/1.


\newpage
\appendix
\section{Linear mode calculations}\label{app:lin}
Assuming that the background velocity ${\bf u}_0$ and density $\rho_0$ are time-independent in the precessing frame, \autoref{eq:vecl1} and \autoref{eq:vecl2} can be written as
\begin{align}
    \partial_t v^i
    +u_0^j\partial_j v^i
    +v^j\partial_j u_0^i
    +2\Gamma_{kj}^iu_0^kv^j
    +2Jg^{ij}\epsilon_{jkl}\Omega_p^kv^l
    &=-c_s^2g^{ij}\partial_jh,
\\
    J\rho_0\partial_t h
    +\partial_j(J\rho_0v^j)
    +\partial_j(J\rho_0u_0^jh)
    &=0.
\end{align}
Here $J=\sqrt{\text{det}g_{ij}}$ is the Jacobian for a general curvilinear coordinate system, $g_{ij}$ is the metric tensor, $g^{ij}$ is the inverse metric tensor, $\Gamma^i_{jk}$ are Christoffel symbols, and paired upper/lower indices (which indicate contravariance/covariance) imply summation. 

We adopt orbital coordinates $(\lambda,\phi,z)$ \citep{Ogilvie2001}, which combine the cylindrical polar coordinates $\phi,z$ with a quasi-radial coordinate defined by the orbital semilatus rectum $\lambda=a(1-e^2)$. \citet{Ogilvie2014} catalogues geometric factors associated with this coordinate system. Here we note that $\lambda,$ $\phi$, and cylindrical radius $r$ are related by $R=\lambda[1 + e\cos(\phi - \varpi)]^{-1}$, where $\varpi$ is the longitude of pericentre. The background velocity field ${\bf u}_0$ in a frame precessing with an untwisted eccentric disk around a reference mass $M$ is additionally given by
\begin{align}\label{eq:u0}
    {\bf u}_0
    &=\sqrt{\frac{GM}{\lambda}}
    \left[
        e\sin\phi\hat{\bf r}
        +(1 + e\cos\phi)\hat{\boldsymbol{\phi}}
    \right]
    =\sqrt{\frac{GM}{\lambda^3}}
    (1 + e\cos\phi)^2{\bf E}_\phi
    =\Omega{\bf E}_\phi,
\end{align}
where ${\bf E}_i$ are the natural basis vectors associated with orbital coordinates (i.e., $u_0^i=\Omega\delta^i_\phi,$ where $\delta_j^i$ is the Kronecker delta). Meanwhile, the background density for an untwisted disk takes the form \citep{Barker2016}
\begin{equation}
    \rho_0
    =\Sigma_b(\lambda)
    \frac{(1-e^2)^{3/2}(1+e\cos\phi)}
    {1 +(e -\lambda e')\cos\phi 
    },
\end{equation}
where $\Sigma_b(\lambda)$ is an arbitrary function of semilatus rectum determined by the profile of mass density.

The independence of the background state from $z$ and $t$ permits the assumption ${\bf v},h\propto\exp[\text{i}(k_zz-\omega t)]$ for a vertical wavenumber $k_z$ and (precessing-frame) frequency $\omega$. Trading $v^\lambda$ for $\tilde{v}^\lambda=\text{i}v^\lambda$ and $v^\phi$ for $w^\phi=\lambda v^\phi$, the linearized equations can then be expressed in component form as
\begin{align}\label{eq:proj1}
    \omega \tilde{v}^\lambda 
    &=-\text{i}(
        \Omega\partial_\phi
        +2\Omega\Gamma^\lambda_{\lambda\phi}
        +2g^{\lambda \phi}J\omega_P
    )\tilde{v}^\lambda
    +2\lambda^{-1}(
        \Omega\Gamma^\lambda_{\phi\phi}
        -g^{\lambda \lambda}J\omega_P
    )w^\phi
    +c_s^2(g^{\lambda\lambda}\partial_\lambda 
    +g^{\lambda\phi}\partial_\phi)h,
\\
    \omega w^\phi
    &=-\lambda (
        \partial_\lambda \Omega 
        +2\Omega\Gamma^\phi_{\lambda\phi}
        +2g^{\phi\phi}J\omega_P
    )\tilde{v}^\lambda
    -\text{i}(
        \Omega\partial_\phi
        +\partial_\phi \Omega 
        +2\Omega\Gamma^\phi_{\phi\phi}
        -2g^{\phi\lambda}J\omega_P
    )w^\phi
    -\text{i}c_s^2\lambda(
        g^{\phi\lambda}\partial_\lambda 
        +g^{\phi\phi}\partial_\phi
    )h,
\\
    \omega v^z
    &=-\text{i}\Omega\partial_\phi v^z
    +k_zc_s^2 h,
\\\label{eq:proj4}
    \omega h\ 
    &=-[
        \partial_\lambda 
        +\partial_\lambda\ln(J\Sigma)
    ]\tilde{v}^\lambda
    -\text{i}\lambda^{-1}[
        \partial_\phi 
        +\partial_\phi\ln(J\Sigma)
    ] w^\phi
    +k_zv^z
    -\text{i}\Omega\partial_\phi h.
\end{align}
Here we have used the fact that $\partial_\phi(J\rho_0\Omega)=0$, which in turn follows from noting that $\rho_0$ and ${\bf u}_0=\Omega{\bf E}_\phi$ constitute a steady solution to the un-perturbed continuity equation in the precessing frame.

\subsection{Numerical calculations}
Inserting the Fourier expansions
\begin{align}
    v^j(\lambda,\phi,z,t)&
    =\sum_{m=-\infty}^\infty v_m^j(\lambda)
    \exp[\text{i}(m\phi+k_zz-\omega t)],
    \\
    h(\lambda,\phi,z,t)&
    =\sum_{m=-\infty}^\infty h_m(\lambda)
    \exp[\text{i}(m\phi+k_zz-\omega t)],
\end{align}
and integrating over $(2\pi)^{-1}\int_0^{2\pi}e^{-\text{i}n\phi}\cdot\text{d}\phi$ for an arbitrary $n$ produces the system of coupled ordinary differential equations 
\begin{align}    
    \omega \tilde{v}^\lambda_n 
    &=a_1^{nm}\tilde{v}^\lambda_m
    +a_2^{nm}w^\phi_m
    +(a_3^{nm}\partial_\lambda + a_4^{nm})h_m,
\\
    \omega w^\phi_n
    &=b_1^{nm}\tilde{v}^\lambda_m
    +b_2^{nm}w^\phi_m
    +(b_3^{nm}\partial_\lambda + b_4^{nm})h_m,
\\
    \omega v^z_n
    &=\Omega^{nm}v^z_m
    +k_zc_s^2 h_n,
\\ 
    \omega h_n 
    &=-\partial_\lambda \tilde{v}^\lambda_n
    +d_1^{nm}\tilde{v}^\lambda_m
    +n\lambda^{-1}w^\phi_n
    +d_2^{nm}w^\phi_m
    +k_zv^z_n
    +\Omega^{nm}h_m,
\end{align}

where paired upper and lower indices indicate summation over all $m\in[-\infty,\infty]$ (but do not indicate co/contravariance), and the ($\lambda$-dependent) coefficients are given by the integrals
\begin{align}
    a_1^{nm}
    &=\int_0^{2\pi}[
        m\Omega
        -2\text{i}(\Omega\Gamma^\lambda_{\lambda\phi}
            +g^{\lambda \phi}J\omega_P
        )
    ]\exp[-\text{i}(n-m)\phi]\text{d}\phi,
\\
    a_2^{nm}
    &=\int_0^{2\pi}
    2\lambda^{-1}(
        \Omega\Gamma^\lambda_{\phi\phi}
        -g^{\lambda \lambda}J\omega_P
    )
    \exp[-\text{i}(n-m)\phi]\text{d}\phi,
\\
    a_3^{nm}
    &=\int_0^{2\pi}c_s^2g^{\lambda\lambda}\exp[-\text{i}(n-m)\phi]\text{d}\phi,
\\
    a_4^{nm}
    &=\int_0^{2\pi}\text{i}mc_s^2g^{\lambda\phi}\exp[-\text{i}(n-m)\phi]\text{d}\phi,
\\
    b_1^{nm}
    &=\int_0^{2\pi}
    -\lambda (
        \partial_\lambda \Omega 
        +2\Omega\Gamma^\phi_{\lambda\phi}
        +2g^{\phi\phi}J\omega_P
    )\exp[-\text{i}(n-m)\phi]\text{d}\phi,
\\
    b_2^{nm}
    &=\int_0^{2\pi}
    [
        m\Omega
        -\text{i}\partial_\phi \Omega 
        -2\text{i}(
        \Omega\Gamma^\phi_{\phi\phi}
        -g^{\phi\lambda}J\omega_P
        )
    ]\exp[-\text{i}(n-m)\phi]\text{d}\phi,
\\
    b_3^{nm}
    &=\int_0^{2\pi}
    -\text{i}c_s^2\lambda g^{\phi\lambda}
    \exp[-\text{i}(n-m)\phi]\text{d}\phi,
\\
    b_4^{nm}
    &=\int_0^{2\pi}mc_s^2\lambda g^{\phi\phi}
    \exp[-\text{i}(n-m)\phi]\text{d}\phi,
\\
    \Omega^{nm}
    &=\int_0^{2\pi}m\Omega \exp[-\text{i}(n-m)\phi]\text{d}\phi,
\\
    d_1^{nm}
    &=\int_0^{2\pi}-\partial_\lambda\ln(J\Sigma)\exp[-\text{i}(n-m)\phi]\text{d}\phi,
\\
    d_2^{nm}
    &=\int_0^{2\pi}-\text{i}\lambda^{-1}\partial_\phi\ln(J\Sigma)
    \exp[-\text{i}(n-m)\phi]\text{d}\phi.
\end{align}
All of these projection integrals are of the form $\int_0^{2\pi }f \exp[-\text{i}(n-m)\phi]\text{d}\phi$ for functions $f=f(\lambda,\phi)$. Since the functions $f$ are in this case periodic on $\phi\in[0,2\pi]$, they can be computed efficiently and accurately with fast fourier transforms. Additionally, our choice of variable ensures that each of these integrals is real-valued. 

\begin{figure}
    \centering
    \includegraphics[width=\linewidth]{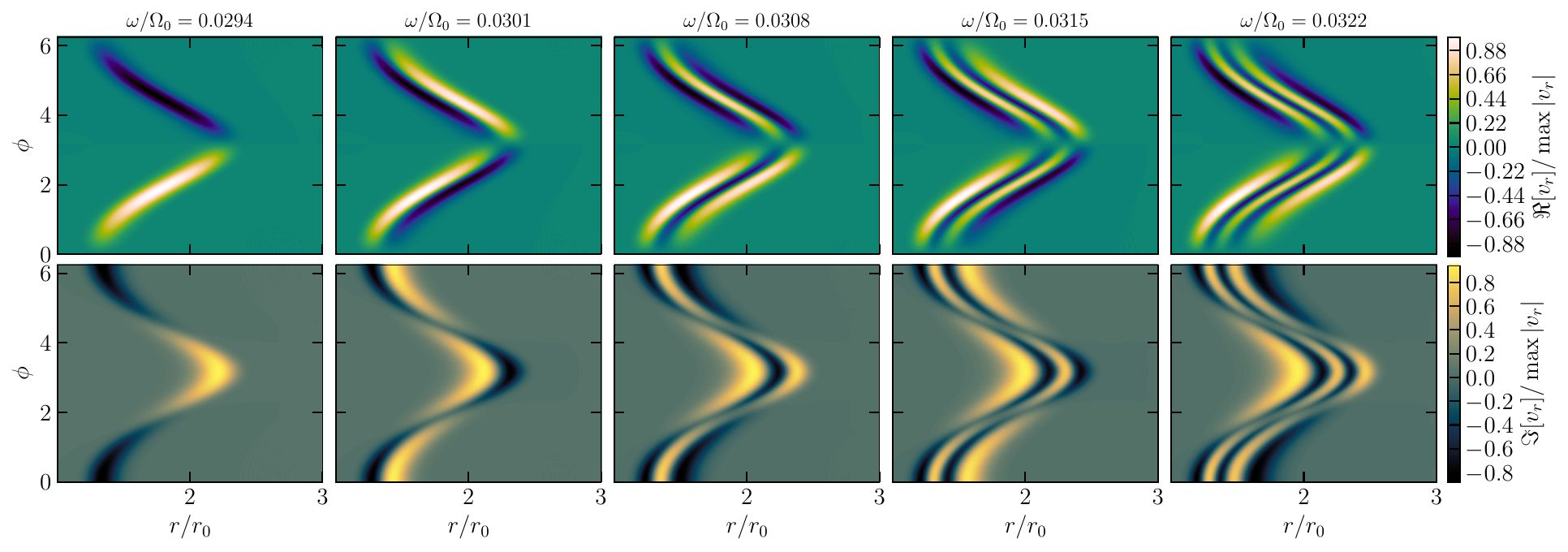}
    \caption{Real (top) and imaginary (bottom) parts of radial velocity perturbations computed for the spectrum of vertically structured eccentric modes computed by solving Equations \eqref{eq:proj1}-\eqref{eq:proj4} with $k_z=4\pi/H_0$ and a background eccentric disk with $\max[e]=0.25, r_1/r_0=3, c_s=0.05r_0\Omega_0.$ The leftmost panel shows the same mode plotted in the righthand panels of \autoref{fig:A25r3Dp2H_globmode}.}
    \label{fig:kz4pi_spec}
\end{figure}

\begin{figure}
    \centering
    \includegraphics[width=\linewidth]{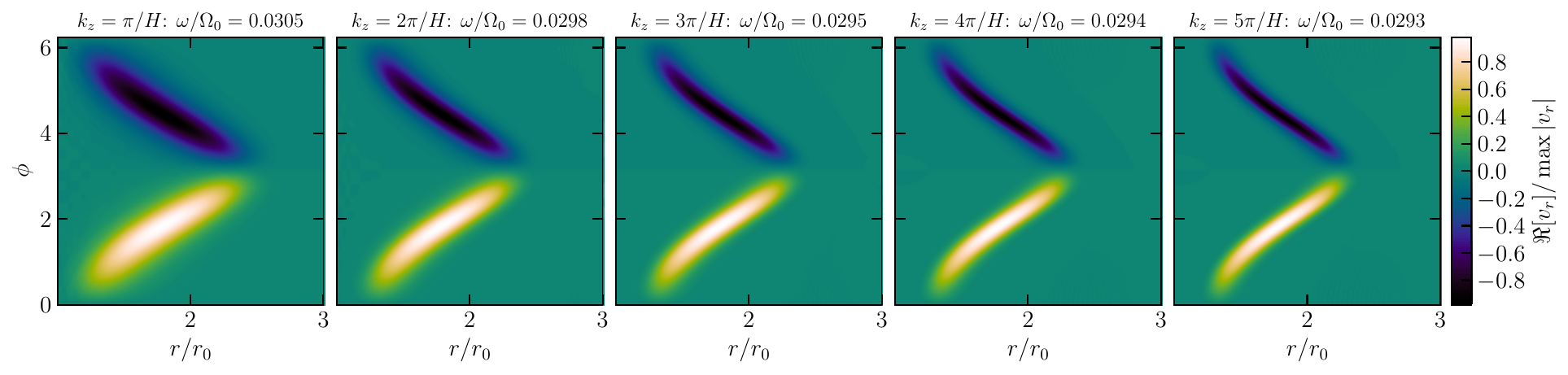}
    \caption{Radial velocity perturbations computed with increasing $k_z$ for the mode shown in the leftmost panel of \autoref{fig:kz4pi_spec}. Increasing $k_z$ narrows the confinement of the mode to the most eccentric streamlines in the disk.}
    \label{fig:kzvar_n1}
\end{figure}

Truncating the Fourier expansions at $m=-M$ and $M$, Equations \eqref{eq:proj1}-\eqref{eq:proj4} constitute a system of $4(2M+1)$ ordinary differential equations in $\lambda$. Adopting the very simple boundary conditions that $\tilde{v}^\lambda_m=0$ at the inner and outer radial boundaries for each $m$, we solve these coupled equations using pseudospectral collocation on a Gauss-Lobatto grid \citep[e.g.,][]{Boyd2001}. Test calculations using the spectral code Dedalus \citep{Burns2020} yielded identical results. \autoref{fig:kz4pi_spec} shows the radial velocity perturbations computed for a spectrum of vertically structured eccentric modes with fixed $k_z$, while \autoref{fig:kzvar_n1} illustrates the effect of changing $k_z$ on a single mode. 

\bibliographystyle{mnras}
\bibliography{manuscript} 

\begin{thebibliography}{}
\makeatletter
\relax
\def\mn@urlcharsother{\let\do\@makeother \do\$\do\&\do\#\do\^\do\_\do\%\do\~}
\def\mn@doi{\begingroup\mn@urlcharsother \@ifnextchar [ {\mn@doi@} {\mn@doi@[]}}
\def\mn@doi@[#1]#2{\def\@tempa{#1}\ifx\@tempa\@empty \href {http://dx.doi.org/#2} {doi:#2}\else \href {http://dx.doi.org/#2} {#1}\fi \endgroup}
\def\mn@eprint#1#2{\mn@eprint@#1:#2::\@nil}
\def\mn@eprint@arXiv#1{\href {http://arxiv.org/abs/#1} {{\tt arXiv:#1}}}
\def\mn@eprint@dblp#1{\href {http://dblp.uni-trier.de/rec/bibtex/#1.xml} {dblp:#1}}
\def\mn@eprint@#1:#2:#3:#4\@nil{\def\@tempa {#1}\def\@tempb {#2}\def\@tempc {#3}\ifx \@tempc \@empty \let \@tempc \@tempb \let \@tempb \@tempa \fi \ifx \@tempb \@empty \def\@tempb {arXiv}\fi \@ifundefined {mn@eprint@\@tempb}{\@tempb:\@tempc}{\expandafter \expandafter \csname mn@eprint@\@tempb\endcsname \expandafter{\@tempc}}}

\bibitem[\protect\citeauthoryear{{Barker} \& {Ogilvie}}{{Barker} \& {Ogilvie}}{2014}]{Barker2014}
{Barker} A.~J.,  {Ogilvie} G.~I.,  2014, \mn@doi [\mnras] {10.1093/mnras/stu1939}, \href {https://ui.adsabs.harvard.edu/abs/2014MNRAS.445.2637B} {445, 2637}

\bibitem[\protect\citeauthoryear{{Barker} \& {Ogilvie}}{{Barker} \& {Ogilvie}}{2016}]{Barker2016}
{Barker} A.~J.,  {Ogilvie} G.~I.,  2016, \mn@doi [\mnras] {10.1093/mnras/stw580}, \href {https://ui.adsabs.harvard.edu/abs/2016MNRAS.458.3739B} {458, 3739}

\bibitem[\protect\citeauthoryear{Boyd}{Boyd}{2001}]{Boyd2001}
Boyd J.~P.,  2001, Chebyshev and Fourier spectral methods, 2 edn.
Dover Publications, Mineola NY

\bibitem[\protect\citeauthoryear{{Burns}, {Vasil}, {Oishi}, {Lecoanet}  \& {Brown}}{{Burns} et~al.}{2020}]{Burns2020}
{Burns} K.~J.,  {Vasil} G.~M.,  {Oishi} J.~S.,  {Lecoanet} D.,   {Brown} B.~P.,  2020, \mn@doi [Physical Review Research] {10.1103/PhysRevResearch.2.023068}, \href {https://ui.adsabs.harvard.edu/abs/2020PhRvR...2b3068B} {2, 023068}

\bibitem[\protect\citeauthoryear{{Calzavarini}, {Doering}, {Gibbon}, {Lohse}, {Tanabe}  \& {Toschi}}{{Calzavarini} et~al.}{2006}]{Calzavarini06}
{Calzavarini} E.,  {Doering} C.~R.,  {Gibbon} J.~D.,  {Lohse} D.,  {Tanabe} A.,   {Toschi} F.,  2006, \mn@doi [\pre] {10.1103/PhysRevE.73.035301}, \href {https://ui.adsabs.harvard.edu/abs/2006PhRvE..73c5301C} {73, 035301}

\bibitem[\protect\citeauthoryear{{Chan}, {Krolik}  \& {Piran}}{{Chan} et~al.}{2018}]{Chan2018}
{Chan} C.-H.,  {Krolik} J.~H.,   {Piran} T.,  2018, \mn@doi [\apj] {10.3847/1538-4357/aab15c}, \href {https://ui.adsabs.harvard.edu/abs/2018ApJ...856...12C} {856, 12}

\bibitem[\protect\citeauthoryear{{Chan}, {Piran}  \& {Krolik}}{{Chan} et~al.}{2022}]{Chan2022}
{Chan} C.-H.,  {Piran} T.,   {Krolik} J.~H.,  2022, \mn@doi [\apj] {10.3847/1538-4357/ac68f3}, \href {https://ui.adsabs.harvard.edu/abs/2022ApJ...933...81C} {933, 81}

\bibitem[\protect\citeauthoryear{{Chan}, {Piran}  \& {Krolik}}{{Chan} et~al.}{2024}]{Chan2024}
{Chan} C.-H.,  {Piran} T.,   {Krolik} J.~H.,  2024, \mn@doi [\apj] {10.3847/1538-4357/ad5d5a}, \href {https://ui.adsabs.harvard.edu/abs/2024ApJ...973..103C} {973, 103}

\bibitem[\protect\citeauthoryear{{Cui} \& {Latter}}{{Cui} \& {Latter}}{2022}]{Cui2022}
{Cui} C.,  {Latter} H.~N.,  2022, \mn@doi [\mnras] {10.1093/mnras/stac279}, \href {https://ui.adsabs.harvard.edu/abs/2022MNRAS.512.1639C} {512, 1639}

\bibitem[\protect\citeauthoryear{{Dai}, {Facchini}, {Clarke}  \& {Haworth}}{{Dai} et~al.}{2015}]{Dai2015}
{Dai} F.,  {Facchini} S.,  {Clarke} C.~J.,   {Haworth} T.~J.,  2015, \mn@doi [\mnras] {10.1093/mnras/stv403}, \href {https://ui.adsabs.harvard.edu/abs/2015MNRAS.449.1996D} {449, 1996}

\bibitem[\protect\citeauthoryear{{Dewberry}, {Latter}, {Ogilvie}  \& {Fromang}}{{Dewberry} et~al.}{2020a}]{Dewberry2020a}
{Dewberry} J.~W.,  {Latter} H.~N.,  {Ogilvie} G.~I.,   {Fromang} S.,  2020a, \mn@doi [\mnras] {10.1093/mnras/staa1897}, \href {https://ui.adsabs.harvard.edu/abs/2020MNRAS.497..435D} {497, 435}

\bibitem[\protect\citeauthoryear{{Dewberry}, {Latter}, {Ogilvie}  \& {Fromang}}{{Dewberry} et~al.}{2020b}]{Dewberry2020b}
{Dewberry} J.~W.,  {Latter} H.~N.,  {Ogilvie} G.~I.,   {Fromang} S.,  2020b, \mn@doi [\mnras] {10.1093/mnras/staa1898}, \href {https://ui.adsabs.harvard.edu/abs/2020MNRAS.497..451D} {497, 451}

\bibitem[\protect\citeauthoryear{{Dziembowski} \& {Krolikowska}}{{Dziembowski} \& {Krolikowska}}{1985}]{Dziembowski1985}
{Dziembowski} W.,  {Krolikowska} M.,  1985, \actaa, \href {https://ui.adsabs.harvard.edu/abs/1985AcA....35....5D} {35, 5}

\bibitem[\protect\citeauthoryear{{Fairbairn} \& {Ogilvie}}{{Fairbairn} \& {Ogilvie}}{2023}]{Fairbairn2023}
{Fairbairn} C.~W.,  {Ogilvie} G.~I.,  2023, \mn@doi [\mnras] {10.1093/mnras/stad211}, \href {https://ui.adsabs.harvard.edu/abs/2023MNRAS.520.1022F} {520, 1022}

\bibitem[\protect\citeauthoryear{{Faure}, {Fromang}  \& {Latter}}{{Faure} et~al.}{2014}]{Faure2014}
{Faure} J.,  {Fromang} S.,   {Latter} H.,  2014, \mn@doi [\aap] {10.1051/0004-6361/201321911}, \href {https://ui.adsabs.harvard.edu/abs/2014A&A...564A..22F} {564, A22}

\bibitem[\protect\citeauthoryear{{Ferreira} \& {Ogilvie}}{{Ferreira} \& {Ogilvie}}{2008}]{Ferreira2008}
{Ferreira} B.~T.,  {Ogilvie} G.~I.,  2008, \mn@doi [\mnras] {10.1111/j.1365-2966.2008.13207.x}, \href {https://ui.adsabs.harvard.edu/abs/2008MNRAS.386.2297F} {386, 2297}

\bibitem[\protect\citeauthoryear{{Ferreira} \& {Ogilvie}}{{Ferreira} \& {Ogilvie}}{2009}]{Ferreira2009}
{Ferreira} B.~T.,  {Ogilvie} G.~I.,  2009, \mn@doi [\mnras] {10.1111/j.1365-2966.2008.14070.x}, \href {https://ui.adsabs.harvard.edu/abs/2009MNRAS.392..428F} {392, 428}

\bibitem[\protect\citeauthoryear{{Fromang}, {Hennebelle}  \& {Teyssier}}{{Fromang} et~al.}{2006}]{Fromang2006}
{Fromang} S.,  {Hennebelle} P.,   {Teyssier} R.,  2006, \mn@doi [\aap] {10.1051/0004-6361:20065371}, \href {https://ui.adsabs.harvard.edu/abs/2006A&A...457..371F} {457, 371}

\bibitem[\protect\citeauthoryear{{Godon}}{{Godon}}{1998}]{Godon1998}
{Godon} P.,  1998, \mn@doi [\apj] {10.1086/305887}, \href {https://ui.adsabs.harvard.edu/abs/1998ApJ...502..382G} {502, 382}

\bibitem[\protect\citeauthoryear{{Goldreich} \& {Tremaine}}{{Goldreich} \& {Tremaine}}{1981}]{Goldreich1981}
{Goldreich} P.,  {Tremaine} S.,  1981, \mn@doi [\apj] {10.1086/158671}, \href {https://ui.adsabs.harvard.edu/abs/1981ApJ...243.1062G} {243, 1062}

\bibitem[\protect\citeauthoryear{{Goldreich}, {Goodman}  \& {Narayan}}{{Goldreich} et~al.}{1986}]{Goldreich1986}
{Goldreich} P.,  {Goodman} J.,   {Narayan} R.,  1986, \mn@doi [\mnras] {10.1093/mnras/221.2.339}, \href {https://ui.adsabs.harvard.edu/abs/1986MNRAS.221..339G} {221, 339}

\bibitem[\protect\citeauthoryear{{Goodchild} \& {Ogilvie}}{{Goodchild} \& {Ogilvie}}{2006}]{Goodchild2006}
{Goodchild} S.,  {Ogilvie} G.,  2006, \mn@doi [\mnras] {10.1111/j.1365-2966.2006.10197.x}, \href {https://ui.adsabs.harvard.edu/abs/2006MNRAS.368.1123G} {368, 1123}

\bibitem[\protect\citeauthoryear{{Goodman}}{{Goodman}}{1993}]{Goodman1993}
{Goodman} J.,  1993, \mn@doi [\apj] {10.1086/172472}, \href {https://ui.adsabs.harvard.edu/abs/1993ApJ...406..596G} {406, 596}

\bibitem[\protect\citeauthoryear{{Guillochon}, {Manukian}  \& {Ramirez-Ruiz}}{{Guillochon} et~al.}{2014}]{Guillochon2014}
{Guillochon} J.,  {Manukian} H.,   {Ramirez-Ruiz} E.,  2014, \mn@doi [\apj] {10.1088/0004-637X/783/1/23}, \href {https://ui.adsabs.harvard.edu/abs/2014ApJ...783...23G} {783, 23}

\bibitem[\protect\citeauthoryear{{Gurzadian} \& {Ozernoi}}{{Gurzadian} \& {Ozernoi}}{1979}]{Gurzadian1979}
{Gurzadian} V.~G.,  {Ozernoi} L.~M.,  1979, \mn@doi [\nat] {10.1038/280214a0}, \href {https://ui.adsabs.harvard.edu/abs/1979Natur.280..214G} {280, 214}

\bibitem[\protect\citeauthoryear{{Haswell}, {King}, {Murray}  \& {Charles}}{{Haswell} et~al.}{2001}]{Haswell2001}
{Haswell} C.~A.,  {King} A.~R.,  {Murray} J.~R.,   {Charles} P.~A.,  2001, \mn@doi [\mnras] {10.1046/j.1365-8711.2001.04034.x}, \href {https://ui.adsabs.harvard.edu/abs/2001MNRAS.321..475H} {321, 475}

\bibitem[\protect\citeauthoryear{{Held} \& {Ogilvie}}{{Held} \& {Ogilvie}}{2024}]{Held2024}
{Held} L.~E.,  {Ogilvie} G.~I.,  2024, \mn@doi [\mnras] {10.1093/mnras/stae2487}, \href {https://ui.adsabs.harvard.edu/abs/2024MNRAS.535.3108H} {535, 3108}

\bibitem[\protect\citeauthoryear{{Hirose} \& {Osaki}}{{Hirose} \& {Osaki}}{1993}]{Hirose1993}
{Hirose} M.,  {Osaki} Y.,  1993, \pasj, \href {https://ui.adsabs.harvard.edu/abs/1993PASJ...45..595H} {45, 595}

\bibitem[\protect\citeauthoryear{{Holoien} et~al.,}{{Holoien} et~al.}{2019}]{Holoien2019}
{Holoien} T.~W.~S.,  et~al., 2019, \mn@doi [\apj] {10.3847/1538-4357/ab2ae1}, \href {https://ui.adsabs.harvard.edu/abs/2019ApJ...880..120H} {880, 120}

\bibitem[\protect\citeauthoryear{{Kato}}{{Kato}}{2004}]{Kato2004}
{Kato} S.,  2004, \mn@doi [\pasj] {10.1093/pasj/56.5.905}, \href {https://ui.adsabs.harvard.edu/abs/2004PASJ...56..905K} {56, 905}

\bibitem[\protect\citeauthoryear{{Lai} \& {Mu{\~n}oz}}{{Lai} \& {Mu{\~n}oz}}{2023}]{Munoz2023}
{Lai} D.,  {Mu{\~n}oz} D.~J.,  2023, \mn@doi [\araa] {10.1146/annurev-astro-052622-022933}, \href {https://ui.adsabs.harvard.edu/abs/2023ARA&A..61..517L} {61, 517}

\bibitem[\protect\citeauthoryear{{Landau} \& {Lifshitz}}{{Landau} \& {Lifshitz}}{1969}]{Landau1969}
{Landau} L.~D.,  {Lifshitz} E.~M.,  1969, {Mechanics}.
Pergamon Press

\bibitem[\protect\citeauthoryear{{Latter} \& {Balbus}}{{Latter} \& {Balbus}}{2009}]{Latter2009}
{Latter} H.~N.,  {Balbus} S.~A.,  2009, \mn@doi [\mnras] {10.1111/j.1365-2966.2009.15350.x}, \href {https://ui.adsabs.harvard.edu/abs/2009MNRAS.399.1058L} {399, 1058}

\bibitem[\protect\citeauthoryear{{Li}, {Goodman}  \& {Narayan}}{{Li} et~al.}{2003}]{Li2003}
{Li} L.-X.,  {Goodman} J.,   {Narayan} R.,  2003, \mn@doi [\apj] {10.1086/376695}, \href {https://ui.adsabs.harvard.edu/abs/2003ApJ...593..980L} {593, 980}

\bibitem[\protect\citeauthoryear{{Lohse} \& {Shishkina}}{{Lohse} \& {Shishkina}}{2024}]{Lohse24}
{Lohse} D.,  {Shishkina} O.,  2024, \mn@doi [Reviews of Modern Physics] {10.1103/RevModPhys.96.035001}, \href {https://ui.adsabs.harvard.edu/abs/2024RvMP...96c5001L} {96, 035001}

\bibitem[\protect\citeauthoryear{{Lubow}}{{Lubow}}{1991a}]{Lubow1991a}
{Lubow} S.~H.,  1991a, \mn@doi [\apj] {10.1086/170647}, \href {https://ui.adsabs.harvard.edu/abs/1991ApJ...381..259L} {381, 259}

\bibitem[\protect\citeauthoryear{{Lubow}}{{Lubow}}{1991b}]{Lubow1991b}
{Lubow} S.~H.,  1991b, \mn@doi [\apj] {10.1086/170648}, \href {https://ui.adsabs.harvard.edu/abs/1991ApJ...381..268L} {381, 268}

\bibitem[\protect\citeauthoryear{{Lynch} \& {Dewberry}}{{Lynch} \& {Dewberry}}{2023}]{Lynch2023}
{Lynch} E.~M.,  {Dewberry} J.~W.,  2023, \mn@doi [\mnras] {10.1093/mnras/stad2678}, \href {https://ui.adsabs.harvard.edu/abs/2023MNRAS.526.2673L} {526, 2673}

\bibitem[\protect\citeauthoryear{{MacFadyen} \& {Milosavljevi{\'c}}}{{MacFadyen} \& {Milosavljevi{\'c}}}{2008}]{MacFadyen2008}
{MacFadyen} A.~I.,  {Milosavljevi{\'c}} M.,  2008, \mn@doi [\apj] {10.1086/523869}, \href {https://ui.adsabs.harvard.edu/abs/2008ApJ...672...83M} {672, 83}

\bibitem[\protect\citeauthoryear{{Mu{\~n}oz} \& {Lithwick}}{{Mu{\~n}oz} \& {Lithwick}}{2020}]{Munoz2020}
{Mu{\~n}oz} D.~J.,  {Lithwick} Y.,  2020, \mn@doi [\apj] {10.3847/1538-4357/abc74c}, \href {https://ui.adsabs.harvard.edu/abs/2020ApJ...905..106M} {905, 106}

\bibitem[\protect\citeauthoryear{{Narayan}, {Goldreich}  \& {Goodman}}{{Narayan} et~al.}{1987}]{Narayan1987}
{Narayan} R.,  {Goldreich} P.,   {Goodman} J.,  1987, \mn@doi [\mnras] {10.1093/mnras/228.1.1}, \href {https://ui.adsabs.harvard.edu/abs/1987MNRAS.228....1N} {228, 1}

\bibitem[\protect\citeauthoryear{{Ogilvie}}{{Ogilvie}}{2001}]{Ogilvie2001}
{Ogilvie} G.~I.,  2001, \mn@doi [\mnras] {10.1046/j.1365-8711.2001.04416.x}, \href {https://ui.adsabs.harvard.edu/abs/2001MNRAS.325..231O} {325, 231}

\bibitem[\protect\citeauthoryear{{Ogilvie}}{{Ogilvie}}{2008}]{Ogilvie2008}
{Ogilvie} G.~I.,  2008, \mn@doi [\mnras] {10.1111/j.1365-2966.2008.13484.x}, \href {https://ui.adsabs.harvard.edu/abs/2008MNRAS.388.1372O} {388, 1372}

\bibitem[\protect\citeauthoryear{{Ogilvie} \& {Barker}}{{Ogilvie} \& {Barker}}{2014}]{Ogilvie2014}
{Ogilvie} G.~I.,  {Barker} A.~J.,  2014, \mn@doi [\mnras] {10.1093/mnras/stu1795}, \href {https://ui.adsabs.harvard.edu/abs/2014MNRAS.445.2621O} {445, 2621}

\bibitem[\protect\citeauthoryear{{Ogilvie} \& {Latter}}{{Ogilvie} \& {Latter}}{2013}]{Ogilvie2013}
{Ogilvie} G.~I.,  {Latter} H.~N.,  2013, \mn@doi [\mnras] {10.1093/mnras/stt917}, \href {https://ui.adsabs.harvard.edu/abs/2013MNRAS.433.2420O} {433, 2420}

\bibitem[\protect\citeauthoryear{{Ogilvie} \& {Lynch}}{{Ogilvie} \& {Lynch}}{2019}]{Ogilvie2019}
{Ogilvie} G.~I.,  {Lynch} E.~M.,  2019, \mn@doi [\mnras] {10.1093/mnras/sty3436}, \href {https://ui.adsabs.harvard.edu/abs/2019MNRAS.483.4453O} {483, 4453}

\bibitem[\protect\citeauthoryear{{Ogilvie}, {Latter}  \& {Lesur}}{{Ogilvie} et~al.}{2025}]{Ogilvie2025}
{Ogilvie} G.~I.,  {Latter} H.~N.,   {Lesur} G.,  2025, \mn@doi [\mnras] {10.1093/mnras/staf154}, \href {https://ui.adsabs.harvard.edu/abs/2025MNRAS.537.3349O} {537, 3349}

\bibitem[\protect\citeauthoryear{{Okazaki}}{{Okazaki}}{1991}]{Okazaki1991}
{Okazaki} A.~T.,  1991, \pasj, \href {https://ui.adsabs.harvard.edu/abs/1991PASJ...43...75O} {43, 75}

\bibitem[\protect\citeauthoryear{{Ostriker}, {Shu}  \& {Adams}}{{Ostriker} et~al.}{1992}]{Ostriker1992}
{Ostriker} E.~C.,  {Shu} F.~H.,   {Adams} F.~C.,  1992, \mn@doi [\apj] {10.1086/171916}, \href {https://ui.adsabs.harvard.edu/abs/1992ApJ...399..192O} {399, 192}

\bibitem[\protect\citeauthoryear{{Papaloizou}}{{Papaloizou}}{2002}]{Papaloizou2002}
{Papaloizou} J.~C.~B.,  2002, \mn@doi [\aap] {10.1051/0004-6361:20020490}, \href {https://ui.adsabs.harvard.edu/abs/2002A&A...388..615P} {388, 615}

\bibitem[\protect\citeauthoryear{{Papaloizou}}{{Papaloizou}}{2005a}]{Papaloizou2005a}
{Papaloizou} J.~C.~B.,  2005a, \mn@doi [\aap] {10.1051/0004-6361:20041947}, \href {https://ui.adsabs.harvard.edu/abs/2005A&A...432..743P} {432, 743}

\bibitem[\protect\citeauthoryear{{Papaloizou}}{{Papaloizou}}{2005b}]{Papaloizou2005b}
{Papaloizou} J.~C.~B.,  2005b, \mn@doi [\aap] {10.1051/0004-6361:20041948}, \href {https://ui.adsabs.harvard.edu/abs/2005A&A...432..757P} {432, 757}

\bibitem[\protect\citeauthoryear{{Papaloizou} \& {Pringle}}{{Papaloizou} \& {Pringle}}{1984}]{Papaloizou1984}
{Papaloizou} J.~C.~B.,  {Pringle} J.~E.,  1984, \mn@doi [\mnras] {10.1093/mnras/208.4.721}, \href {https://ui.adsabs.harvard.edu/abs/1984MNRAS.208..721P} {208, 721}

\bibitem[\protect\citeauthoryear{{Papaloizou}, {Savonije}  \& {Henrichs}}{{Papaloizou} et~al.}{1992}]{Papaloizou1992}
{Papaloizou} J.~C.,  {Savonije} G.~J.,   {Henrichs} H.~F.,  1992, \aap, \href {https://ui.adsabs.harvard.edu/abs/1992A&A...265L..45P} {265, L45}

\bibitem[\protect\citeauthoryear{{Patterson} et~al.,}{{Patterson} et~al.}{2005}]{Patterson2005}
{Patterson} J.,  et~al., 2005, \mn@doi [\pasp] {10.1086/447771}, \href {https://ui.adsabs.harvard.edu/abs/2005PASP..117.1204P} {117, 1204}

\bibitem[\protect\citeauthoryear{{Pierens}, {McNally}  \& {Nelson}}{{Pierens} et~al.}{2020}]{Pierens2020}
{Pierens} A.,  {McNally} C.~P.,   {Nelson} R.~P.,  2020, \mn@doi [\mnras] {10.1093/mnras/staa1550}, \href {https://ui.adsabs.harvard.edu/abs/2020MNRAS.496.2849P} {496, 2849}

\bibitem[\protect\citeauthoryear{{Pierens}, {Nelson}  \& {McNally}}{{Pierens} et~al.}{2021}]{Pierens2021}
{Pierens} A.,  {Nelson} R.~P.,   {McNally} C.~P.,  2021, \mn@doi [\mnras] {10.1093/mnras/stab2853}, \href {https://ui.adsabs.harvard.edu/abs/2021MNRAS.508.4806P} {508, 4806}

\bibitem[\protect\citeauthoryear{{Reg{\'a}ly}, {S{\'a}ndor}, {Dullemond}  \& {Kiss}}{{Reg{\'a}ly} et~al.}{2011}]{Regaly2011}
{Reg{\'a}ly} Z.,  {S{\'a}ndor} Z.,  {Dullemond} C.~P.,   {Kiss} L.~L.,  2011, \mn@doi [\aap] {10.1051/0004-6361/201016152}, \href {https://ui.adsabs.harvard.edu/abs/2011A&A...528A..93R} {528, A93}

\bibitem[\protect\citeauthoryear{{Ryu} \& {Goodman}}{{Ryu} \& {Goodman}}{1994}]{Ryu1994}
{Ryu} D.,  {Goodman} J.,  1994, \mn@doi [\apj] {10.1086/173725}, \href {https://ui.adsabs.harvard.edu/abs/1994ApJ...422..269R} {422, 269}

\bibitem[\protect\citeauthoryear{{Statler}}{{Statler}}{2001}]{Statler2001}
{Statler} T.~S.,  2001, \mn@doi [\aj] {10.1086/323713}, \href {https://ui.adsabs.harvard.edu/abs/2001AJ....122.2257S} {122, 2257}

\bibitem[\protect\citeauthoryear{{Svanberg}, {Cui}  \& {Latter}}{{Svanberg} et~al.}{2022}]{Svanberg2022}
{Svanberg} E.,  {Cui} C.,   {Latter} H.~N.,  2022, \mn@doi [\mnras] {10.1093/mnras/stac1598}, \href {https://ui.adsabs.harvard.edu/abs/2022MNRAS.514.4581S} {514, 4581}

\bibitem[\protect\citeauthoryear{{Teyssandier} \& {Ogilvie}}{{Teyssandier} \& {Ogilvie}}{2016}]{Teyssandier2016}
{Teyssandier} J.,  {Ogilvie} G.~I.,  2016, \mn@doi [\mnras] {10.1093/mnras/stw521}, \href {https://ui.adsabs.harvard.edu/abs/2016MNRAS.458.3221T} {458, 3221}

\bibitem[\protect\citeauthoryear{{Teyssandier} \& {Ogilvie}}{{Teyssandier} \& {Ogilvie}}{2017}]{Teyssandier2017}
{Teyssandier} J.,  {Ogilvie} G.~I.,  2017, \mn@doi [\mnras] {10.1093/mnras/stx426}, \href {https://ui.adsabs.harvard.edu/abs/2017MNRAS.467.4577T} {467, 4577}

\bibitem[\protect\citeauthoryear{{Teyssier}}{{Teyssier}}{2002}]{Teyssier2002}
{Teyssier} R.,  2002, \mn@doi [\aap] {10.1051/0004-6361:20011817}, \href {https://ui.adsabs.harvard.edu/abs/2002A&A...385..337T} {385, 337}

\bibitem[\protect\citeauthoryear{{Tremaine}}{{Tremaine}}{2001}]{Tremaine2001}
{Tremaine} S.,  2001, \mn@doi [\aj] {10.1086/319398}, \href {https://ui.adsabs.harvard.edu/abs/2001AJ....121.1776T} {121, 1776}

\bibitem[\protect\citeauthoryear{{Whitehurst}}{{Whitehurst}}{1988}]{Whitehurst1988}
{Whitehurst} R.,  1988, \mn@doi [\mnras] {10.1093/mnras/232.1.35}, \href {https://ui.adsabs.harvard.edu/abs/1988MNRAS.232...35W} {232, 35}

\bibitem[\protect\citeauthoryear{{Wienkers} \& {Ogilvie}}{{Wienkers} \& {Ogilvie}}{2018}]{Wienkers2018}
{Wienkers} A.~F.,  {Ogilvie} G.~I.,  2018, \mn@doi [\mnras] {10.1093/mnras/sty899}, \href {https://ui.adsabs.harvard.edu/abs/2018MNRAS.477.4838W} {477, 4838}

\bibitem[\protect\citeauthoryear{{Wu} \& {Goldreich}}{{Wu} \& {Goldreich}}{2002}]{Wu2002}
{Wu} Y.,  {Goldreich} P.,  2002, \mn@doi [\apj] {10.1086/324193}, \href {https://ui.adsabs.harvard.edu/abs/2002ApJ...564.1024W} {564, 1024}

\bibitem[\protect\citeauthoryear{{Zurita}, {Durant}, {Torres}, {Shahbaz}, {Casares}  \& {Steeghs}}{{Zurita} et~al.}{2008}]{Zurita2008}
{Zurita} C.,  {Durant} M.,  {Torres} M.~A.~P.,  {Shahbaz} T.,  {Casares} J.,   {Steeghs} D.,  2008, \mn@doi [\apj] {10.1086/588721}, \href {https://ui.adsabs.harvard.edu/abs/2008ApJ...681.1458Z} {681, 1458}

\makeatother
\end{thebibliography}

\end{document}